\input{amssymb.sty}
\documentstyle [aps,prb,preprint,epsf,eqsecnum]{revtex}
\def\boldkappa{\mbox{\boldmath $\kappa$}}
\def\P{\mathit{P}}
\begin{document}

\title{The Isothermal Binodal Curves Near a Critical Endpoint}

\author{Young C. Kim and Michael E. Fisher}
\address{Institute for Physical Science and Technology, \\ University of Maryland, College Park, Maryland 20742}
\author{Marcia C. Barbosa$^{a)}$\footnote[0]{$^{a)}$Current address: Center for Polymer Studies, Center for Computational Science, and Department of Physics, Boston University, Boston, MA 02215 USA}}
\address{Instituto de Fisica, Universidade Federal do Rio Grande do Sul\\ Caixa Postal 15051, 91500 Porto Alegre, RS, Brazil}

\date{\today}

\maketitle
\vspace{-.1in}
\begin{abstract}
Thermodynamics in the vicinity of a critical endpoint with nonclassical exponents $\alpha$, $\beta$, $\gamma$, $\delta$, $\cdots$ is analyzed in terms of density variables (mole fractions, magnetizations, etc.). The shapes of the isothermal binodals or two-phase coexistence curves are found at and near the endpoint for symmetric and nonsymmetric situations. The spectator- (or noncritical)-phase binodal at $T=T_{e}$ is characterized by an exponent $(\delta +1)/\delta$ $(\simeq 1.21)$ with leading corrections of relative order $1/\delta$ $(\simeq 0.21)$, $\theta_{4}/\beta\delta$ $(\simeq 0.34)$ and $1 - (\beta\delta)^{-1}$ $(\simeq 0.36)$; in contrast to classical (van der Waals, mean field, $\cdots$) theory, the critical endpoint binodal is singular with leading exponent $(1-\alpha)/\beta$ $(\simeq 2.73)$ and corrections which are elucidated; the remaining, $\lambda$-line binodals also display the `renormalized exponent,' $(1-\alpha)/\beta$ but with more singular corrections. (The numerical values quoted here pertain to $(d=3)$-dimensional-fluid or Ising-type systems.)
\end{abstract}

\pagebreak
\section{Introduction and Overview}
\label{sec1}
At a critical point in a fluid (or other Ising-type or $n=1$) system two distinct phases, say, $\beta$ and $\gamma$ become identical: below $T=T_{c}$ these two phases may coexist for appropriate values of the conjugate ordering field (or chemical potential, etc.) \cite{ref1} $h$; above $T_{c}$ they merge into a single phase, say, $\beta\gamma$. If there is some other field variable, \cite{ref1} say $g$, which may be varied without destroying coexistence, the critical point is drawn out into a lambda line, $T=T_{c}(g)$. A typical situation, which lacks any special symmetry, is shown in Fig.~\ref{fig1}. The lambda line, $\lambda$, delimits the phase boundary surface $h=h_{\rho}(g,T)$ labeled $\rho$, on which $\beta$ and $\gamma$ may coexist.

Now in many instances when $g$ is varied, say decreased, another quite distinct phase, $\alpha$, will be encountered. In this case the lambda line terminates at a {\em critical endpoint}, \cite{ref2} which is labeled $E$ in Fig.~\ref{fig1}. At $E$ the phases $\beta$ and $\gamma$ may undergo criticality in the presence of the coexisting noncritical phase $\alpha$ which may be appropriately termed the {\em spectator-phase}. \cite{ref2,ref3} The surface bounding the spectator-phase in the $(g,T,h)$ or field space is labeled $\sigma$; on it $\alpha$ may coexist with phases $\beta\gamma$, $\beta$, or $\gamma$; on the triple line, $\tau$, where the surface $\rho$ meets the surface $\sigma$, all three phases $\alpha$, $\beta$ and $\gamma$ may coexist. 

In a previous study \cite{ref2} (to be denoted {\bf I}), we  discussed the shape of the spectator-phase boundary surface, $g=g_{\sigma}(T,h)$, in the vicinity of the endpoint at $T=T_{e}$ and, by choice of origin, $h=h_{e}\equiv0$. It was found that the surface is singular at $E$ with functions such as $g_{\tau}(T)$, specifying the triple line, and $g_{\sigma}(T_{e};h)$ displaying nonanalytic behavior described by a variety of critical exponents. \cite{ref2,ref5} When, as is normally so, the lambda line is characterized by nonclassical exponents $\alpha$, for the specific heat, $\beta$, for the order parameter, $\delta$, for the critical isotherm, etc., the spectator-phase boundary exponents can all be expressed \cite{ref2} in terms of $\alpha$, $\beta$, and $\delta$. Beyond that it was shown that various dimensionless ratios constructed from the amplitudes of the phase-boundary singularities should be {\em universal} with values also determined by the nature of the bulk criticality on the lambda line. \cite{ref2,ref5}

These conclusions were based on a phenomenological description of the thermodynamic potentials (or Gibbs' free energies) $G^{\alpha}(g,T,h)$ and $G^{\beta\gamma}(g,T,h)$, for the spectator-phase and for the coexisting and critical phases, respectively. The former was assumed to have a power series expansion in the vicinity of $E$; the latter embodied a full scaling representation of the critical line and its neighborhood. \cite{ref2,ref5}

This formulation neglects the essential singularities expected on the $\sigma$ and $\rho$ boundaries; \cite{ref4} these can, however, be discussed \cite{ref5} but play only a negligible quantitative role. Our general phenomenological treatment has been checked by an extensive study of a family of spherical models which exhibit lambda lines and critical endpoints with a range of nonclassical exponents (although $\beta=\frac{1}{2}$ in all cases). \cite{ref6,ref7}

Many experimental examples of critical endpoints are found in multicomponent fluid systems. In the simplest example, which we will particularly bear in mind, two chemical species, B and C, mix as fluids in all proportions at high temperatures forming the phase $\beta\gamma$. At lower temperatures, however, they undergo liquid-liquid phase separation, or demixing, producing phases $\beta$ and $\gamma$ rich in B and C, respectively. Up to a constant shift, the field $h$ may then be taken as the chemical potential difference $\mu_{B}-\mu_{C}$. As the pressure, $p$, or the total chemical potential, $\mu_{B}+\mu_{C}$, either of which we may identify with the field $g$, is reduced, a dilute vapor phase, $\alpha$, appears. Fig.~\ref{fig1} then represents a characteristic overall phase diagram. Now in a typical experiment the temperature $T$ is controlled and may be held fixed: corresponding to Fig.~\ref{fig1}, the appropriate isothermal phase diagrams in the $(g,h)$ plane then have the character shown in Fig.~\ref{fig2} for (a) $T<T_{e}$, (b) $T=T_{e}$, and (c) $T>T_{e}$.

However, the chemical potentials $\mu_{B}$ and $\mu_{C}$, or the fields $h$ and $g$, are normally {\em not} under direct experimental control or observation; rather, the conjugate {\em densities}, $\rho_{B}$ and $\rho_{C}$ (or concentrations of B and C) or, equivalently, the densities
 \begin{equation}
  \rho_{1} = -\left.\frac{\partial}{\partial h}G(g,T,h)\right|_{g,T} \hspace{.2in} \mbox{and}\hspace{.2in} \rho_{2} = -\left.\frac{\partial}{\partial g}G(g,T,h)\right|_{T,h},  \label{eq:1.1}
 \end{equation}
are the prime experimental variables. [Note that in the example envisaged with $g=\mu_{B} + \mu_{C}$ one simply has $\rho_{1}=\frac{1}{2}(\rho_{B}-\rho_{C})$ and $\rho_{2} = \frac{1}{2}(\rho_{B}+\rho_{C})$.] In the density plane $(\rho_{1},\rho_{2})$ the phase boundaries $\rho$ and $\sigma$ are represented by two-phase regions bounded by smooth curves, the so-called {\em binodals} or {\em coexistence curves}: see Fig.~\ref{fig3}. The aim of this article is to analyze in detail and generality the {\em shapes} of these isothermal binodal curves in the vicinity of a critical endpoint. Specifically, we will elucidate the nature of the leading and subdominant singularities that appear in the various binodals labeled ${\cal B}_{e}^{\alpha +}$, ${\cal B}_{<}^{\beta}$, etc., in Fig.~\ref{fig3}.

It appears from Fig.~\ref{fig3}, and detailed analysis bears it out, that the binodal curves for $T\geq T_{e}$ meet with a common tangent at the endpoints $E^{\lambda}$ and $E^{\alpha}$ and at the extended triple points $\tilde{\tau}\alpha$ and $\tilde{\tau}\beta\gamma$ (defined by the intersection of $\tilde{\rho}$, the extended phase boundary $\rho$, with the surface $\sigma$ in the $(g,T,h)$ space: see {\bf I}). Of principal concern, then, is the way in which the binodals depart from linearity. {\em Above} $T_{e}$ one expects analytic binodals but the behavior of the curvatures at $\tilde{\tau}\alpha$ and $\tilde{\tau}\beta\gamma$ as $T\rightarrow T_{e}+$ is then of interest. On the other hand, {\em at} $T=T_{e}$ one expects singular behavior at $E^{\lambda}$ and $E^{\alpha}$. Indeed, Borzi, \cite{ref8} stimulated by Widom, \cite{ref9} discussed the {\em noncritical} binodals at the endpoint, namely, ${\cal B}_{e}^{\alpha\pm}$ in Fig.~\ref{fig3}(b), using the simplest possible phenomenological postulate and geometrical arguments (equivalent to van der Waals and other classical theories). He concluded that the degree of tangency was controlled by a $4/3$ power law (in place of a power 2 for a normal analytic tangency).

Later Klinger, \cite{ref9} using a more general phenomenological classical theory, discussed the {\em critical} endpoint binodals, ${\cal B}_{e}^{\beta}$ and ${\cal B}_{e}^{\gamma}$ analytically: see Fig.~\ref{fig3}(b). However, he found no evidence of singular behavior. Beyond that, Klinger confirmed the leading $4/3$ power in the {\em non}critical or spectator binodal and found that the first correction term carries a $5/3$ power.

On general grounds, however, it seems certain that the powers $4/3$ and $5/3$ must result from the reliance on classical theory which entails the critical exponent values $\alpha = 0$, $\beta=\frac{1}{2}$, and $\delta=3$ in place of the appropriate nonclassical values $\alpha \simeq 0.10_{6}$, $\beta \simeq 0.32_{8}$, and $\delta = (2-\alpha)/\beta - 1 \simeq 4.7_{8}$ which characterize the specific heat, coexistence curve, and critical isotherm of real bulk, ($d=3$)-dimensional fluids (or other systems in the Ising universality class). Indeed, Widom has conjectured \cite{ref9} that in general the $4/3$ power should become $(\delta + 1)/\delta$. This reduces to Borzi's result when $\delta = 3$ but yields an exponent value 1.207 for real fluid systems.

Here we confirm Widom's surmise using the full scaling approach developed in {\bf I}. Furthermore we show that Klinger's correction exponent of $5/3$ is replaced, more generally, by {\em three} exponents, namely, $(2-\alpha + \beta)/\beta\delta$, $(2-\alpha+\theta_{4})/\beta\delta$ and $(3-2\alpha-\beta)/\beta\delta$. Here $\theta_{4}$ is the leading correction-to-scaling exponent which has the value $\theta_{4} \simeq 0.54$ for $(d=3)$-dimensional Ising-type systems; \cite{ref10} thus these three exponents have values of about 1.42, 1.55 and 1.57, respectively, for bulk fluids. In the classical limit attained via $d \rightarrow 4-$ one has $\theta_{4} \rightarrow 0$ and the second exponent reduces \cite{ref11} to $4/3$ while the first and third yield Klinger's value $5/3$. However, we also identify further singular exponents that must appear in the expansion of the noncritical binodal at the endpoint.

It transpires, in addition, that, contrary to Klinger's findings, \cite{ref9} the critical binodal is, in general, {\em also singular} with a leading power $(1-\alpha)/\beta \simeq 2.7_{3}$ so that the binodal is much flatter at the endpoint $E^{\lambda}$ than classical theory would predict. Here, and below where appropriate, we suppose $\alpha > 0$ as applies to real fluids. The exponent $(1-\alpha)/\beta$ is, in fact, the same as that long known to characterize isothermal binodals passing through a lambda point (away from any endpoint): see ${\cal B}^{\lambda +}$ and ${\cal B}^{\lambda -}$ in Fig.~\ref{fig3}(a). This behavior which is, of course, reconfirmed by our analysis reflects, in turn, the phenomenon of critical exponent renormalization. \cite{ref12} The correction terms in the critical endpoint binodal are found to carry exponents $(1-\alpha+\theta_{k})/\beta$ with $k=4,\,5,\,\cdots$. When one substitutes the classical values $\alpha = 0$ and $\theta_{k} = \frac{1}{2}(k-4)$ these leading and correction exponents become 2, 3, 4, $\cdots$ which are consistent with Klinger's results and indicative of a fully analytic critical binodal.

The results sketched here, and others for the remaining binodals shown in Fig.~\ref{fig3}, are presented in detail in Sec.~\ref{sec3}. However, it is necessary to point out that Figs.~\ref{fig1}-3 are special in two respects. First, as mentioned, no symmetry with respect to the ordering surface $\rho$ has been supposed: this is quite appropriate for most fluid systems. However, as observed in {\bf I}, \cite{ref2,ref5} there are many other physical systems in which the thermodynamic potentials are unchanged under reflection in the {\em plane} $\rho$: one may then take $h=0$ on $\rho$ and the symmetry becomes invariance under $h \Longleftrightarrow -h$. The conceptually simplest example is an elemental ferromagnet, like nickel or iron, where $h\equiv H$ is the magnetic field and $g\equiv p$, the pressure. Other examples are ferroelectrics, antiferromagnets, order-disorder binary alloys, and liquid helium through its transition to superfluidity; \cite{ref5} however, the binodal curves are not readily accessible experimentally in some of these cases. The corresponding $(g,T,h)$ phase space, the isothermal sections, and the binodal curves for such {\em symmetric critical endpoints} are illustrated in Figs.~\ref{fig4},~\ref{fig5}, and~\ref{fig6}. In fact, symmetric critical endpoints are simpler in a number of respects and will be analyzed first below. Fundamentally we find that the leading singular behavior of the binodals is identical in the symmetric and nonsymmetric cases but the correction terms differ in character: see Sec.~\ref{sec3}.

A second special feature embodied in Figs.~\ref{fig1}-3 is the {\em slope} of the $\lambda$ line which we characterize as {\em negative} in the sense that if, without loss of generality, we (i) take
 \begin{equation}
  g = h = 0, \hspace{.3in} T = T_{e}, \hspace{.3in} \mbox{at $E$},  \label{eq:1.2}
 \end{equation}
{\em and} (ii) suppose that the negative $g$ axis lies in the $\alpha$ or spectator-phase (see Figs. 1 and 4) then we have \cite{ref2}
 \begin{equation}
  \mbox{{\bf A}}:\hspace{.3in} \Lambda_{g} \equiv T_{e}\left(\frac{dT_{c}}{dg}\right)_{e}^{-1} < 0.  \label{eq:1.3}
 \end{equation}
Conversely, as illustrated in Fig.~\ref{fig4}, one must also consider the case of a positively sloping $\lambda$ line with
 \begin{equation}
  \mbox{{\bf B}}:\hspace{.3in} \Lambda_{g} \equiv T_{e}\left(\frac{dT_{c}}{dg}\right)_{e}^{-1} > 0.  \label{eq:1.4}
 \end{equation}
As seen in Figs.~\ref{fig5} and~\ref{fig6}, this produces distinct isothermal phase diagrams and new arrangements of binodal curves: note the additional notation introduced in Fig.~\ref{fig6}.

One might, of course, also wish to consider the borderline cases $\Lambda_{g} = 0$, $\infty$; we will not pursue these but, on the basis of our postulates for the thermodynamic potentials as set out below in Sec.~\ref{sec2}, the necessary analysis presents no further problems of principle.

In summary therefore, we will analyze the binodals for four types of critical endpoints which, using {\bf N} for nonsymmetric and {\bf S} for symmetric, may be labeled {\bf NA} (Figs.~\ref{fig1}-3), {\bf NB}, {\bf SA}, and {\bf SB} (Figs.~\ref{fig4}-6).

In outline the remainder of the paper is as follows: Our basic scaling postulates for the thermodynamic potential $G^{\beta\gamma}(g,T,h)$ are set out in Sec.~\ref{sec2}. They are essentially the same as introduced and discussed critically in {\bf I} but they have been extended significantly as regards the {\em symmetries} of the corrections to scaling and of the nonlinear scaling fields; the notation also differs in a few details. \cite{ref2} The reader prepared to take the postulates on trust \cite{ref7} may proceed directly to Sec.~\ref{sec3} where the shapes of the binodals in the various cases are discussed in detail without reference to Sec.~\ref{sec2}. The analytic derivation of the results, which is straightforward in principle but a little delicate in practice, is presented in Sec.~\ref{sec4}. Explicit formulae for the many amplitudes entering the expressions for the various binodals in Sec.~\ref{sec3} are also given in Sec.~\ref{sec4}. Section~\ref{sec5} summarizes our conclusions briefly.

\section{Thermodynamic Potentials for Endpoints}
\label{sec2}
This section sets out a complete specification of the thermodynamic potential $G(g,T,h)$ in field variables as needed for the general description of critical endpoints. It is the basis for the results described in Sec.~\ref{sec3} but need not be read to understand those results. For convenience we adopt the critical endpoint as origin for the fields $g$ and $h$ as specified in (\ref{eq:1.2}), and also put
 \begin{equation}
  t = (T-T_{e})/T_{e}.  \label{eq:2.1}
 \end{equation}
Thus $g$, $t$, and $h$ measure field deviations from the endpoint $E$ at $(g,t,h) = (0,0,0)$. (In {\bf I} the variables $g$ and $t$ were denoted $\Delta g$ and $\hat{t}$.) For any property $P(g,T,h)$ admitting a power series expansion about $E$ (of indefinitely high order but not necessarily convergent) we utilize, for brevity, the semisystematic subscript notation
 \begin{eqnarray}
  P(g,T,h) & = & P_{e} + P_{1}g + P_{2}t + P_{3}h + P_{4}g^{2} + 2P_{5}gh + 2P_{6}gt \nonumber \\
           &   & \hspace{.21in}+ 2P_{7}ht + P_{8}t^{2} + P_{9}h^{2} + O_{3}(g,t,h),  \label{eq:2.2}
 \end{eqnarray}
where, here and below, $O_{m}(x,y,z)$ denotes a formal expansion in powers $x^{j}y^{k}z^{l}$ with $j+k+l\geq m$. If $P$ is {\em symmetric} under $h \Longleftrightarrow -h$ one has
 \begin{equation}
  P_{3} = P_{5} = P_{7} = 0 \hspace{.5in} \mbox{(to order 3)}.   \label{eq:2.3}
 \end{equation}
Functions satisfying (\ref{eq:2.2}) and (\ref{eq:2.3}) will be said to be {\em noncritical} (as against {\em critical}).

Following {\bf I} we assume that the thermodynamic potential $G^{\alpha}(g,T,h)$ for the spectator-phase, $\alpha$, is {\em noncritical}. Thus one has, e.g., $G^{\alpha}_{7} = \frac{1}{2}[\partial^{2}G^{\alpha}(g,T,h)/\partial h \partial t]_{e}$ and, by virtue of (\ref{eq:1.1}), the endpoint densities in the spectator-phase are simply
 \begin{equation}
  \rho_{1}^{\alpha e} = -G^{\alpha}_{3} \hspace{.3in} \mbox{and} \hspace{.3in} \rho_{2}^{\alpha e} = -G^{\alpha}_{1}.  \label{eq:2.4}
 \end{equation}

To describe the critical phases, $\beta$, $\gamma$ and $\beta\gamma$, we first introduce, again following {\bf I}, the two relevant {\em nonlinear} `thermal' and `ordering' scaling fields, $\tilde{t}(g,T,h)$ and $\tilde{h}(g,T,h)$, which both {\em vanish} on the $\lambda$ line while $\tilde{h}$ also vanishes on the phase boundary $\rho$. For the nonlinear scaling fields we accept the noncritical expansions \cite{ref16}
 \begin{eqnarray}
  \tilde{t} = t & + & q_{0}h + q_{1}g + q_{2}g^{2} + q_{3}gt + q_{4}t^{2} + q_{5}gh \nonumber \\
            & + & q_{6}h^{2} + q_{7}th + O_{3}(g,t,h),    \label{eq:2.5} \\
  \tilde{h} = h & + & r_{-1}t + r_{0}g + r_{1}gh + r_{2}th + r_{3}h^{2} + r_{4}g^{2} \nonumber \\
            & + & r_{5}gt + r_{6}t^{2} + O_{3}(g,t,h),    \label{eq:2.6}
 \end{eqnarray}
which slightly extend those in {\bf I}(4.7), (4,8). It should also be mentioned at this point that {\em pressure-mixing} terms, which have been discovered recently in connection with the Yang-Yang anomaly in fluid systems, \cite{mef:gok,gok:mef:cut} are {\em not} considered here. \cite{mixing} 

In the symmetric case one has, to order 3,
 \begin{equation}
  q_{0} = q_{5} = q_{7} = 0 \hspace{.2in} \mbox{and} \hspace{.2in} r_{j} = 0 \hspace{.2in} \mbox{for} \hspace{.1in} j = -1,\,0,\,3\mbox{-}6.   \label{eq:2.7}
 \end{equation}
Asymptotically, the $\lambda$ line may thus be described by 
 \begin{equation}
  g_{\lambda}(T) = \Lambda_{g}t + \Lambda_{g2}t^{2} + O(t^{3}), \hspace{.4in}
  h_{\lambda}(T) = \Lambda_{h}t + \Lambda_{h2}t^{2} + O(t^{3}),   \label{eq:2.8}
 \end{equation}
where one finds
 \begin{equation}
  \Lambda_{g} = -\frac{1 - q_{0}r_{-1}}{q_{1} - q_{0}r_{0}}, \hspace{.3in} \Lambda_{h} = \frac{r_{0} - q_{1}r_{-1}}{q_{1} - q_{0}r_{0}},   \label{eq:2.9}
 \end{equation}
with similar expressions for $\Lambda_{g2}$, etc. In accord with (\ref{eq:1.3}) and (\ref{eq:1.4}), we assume $\Lambda_{g}$ does not vanish or diverge. In the symmetric case one has $\Lambda_{g} = -1/q_{1}$ and $\Lambda_{h} = \Lambda_{h2} = \cdots = 0$, so that $\Lambda_{0} \equiv q_{1} - q_{0}r_{0} \neq 0$.

Then we need the one {\em relevant} scaled variable
 \begin{equation}
  y(g,t,h) = U\tilde{h}/|\tilde{t}|^{\Delta}, \hspace{.2in} \mbox{with} \hspace{.2in} \Delta = \beta\delta = \beta + \gamma > 1,  \label{eq:2.10}
 \end{equation}
where the exponent relations and inequality are standard. In {\bf I} we took $U = U(g,t,h)$ as a noncritical function; however, with no loss of generality we may take $U$ as a {\em positive constant} since any dependence on $g$, $t$, and $h$ can be absorbed into $\tilde{h}$. Beyond $y$ we need the many {\em irrelevant} scaled variables
 \begin{eqnarray}
  y_{k}(g,t,h) & = & U_{k}(g,t,h)|\tilde{t}|^{\theta_{k}},  \nonumber \\
               &   & \theta_{k+1} \geq \theta_{k} > 0, \hspace{.3in} k = 4,\, 5, \cdots.   \label{eq:2.11}
 \end{eqnarray}
We assume that the associated irrelevant amplitudes $U_{k}$ are noncritical \cite{ref16} with
 \begin{equation}
  U_{k}(g,t,-h) = (-)^{k}U_{k}(g,t,h) \hspace{.3in} \mbox{in case {\bf S}}.  \label{eq:2.12}
 \end{equation}

Now we can write the thermodynamic potential for the critical phase as
 \begin{equation}
  G^{\beta\gamma}(g,T,h) = G^{0}(g,T,h) - Q|\tilde{t}|^{2-\alpha}W_{\pm}(y,y_{4},y_{5},\cdots),  \label{eq:2.13}
 \end{equation}
where the background $G^{0}(g,T,h)$ and the {\em positive} amplitude $Q(g,T,h)$ are {\em noncritical} while the subscript $\pm$ refers to $\tilde{t}\gtrless 0$. Physically, from the relation of $\alpha$ to the specific heat we have $2 - \alpha > 1$ but we further suppose
 \begin{equation}
  (2 - \alpha)/\Delta = (\delta + 1)/\delta > 1,    \label{eq:2.14}
 \end{equation}
as is generally valid both classically and nonclassically. For concreteness and simplicity we will, in addition, focus on $\alpha > 0$ (as appropriate for bulk fluids, etc.).

We also assume, acknowledging the symmetry of the standard universality classes, that the scaling function $W_{\pm}(y, y_{4}, y_{5}, \cdots)$ is both universal and invariant under change of sign of the odd arguments $y$, $y_{5}$, $y_{7}$, $\cdots$. Beyond that we have the expansion
 \begin{eqnarray}
  W_{\pm}(y, y_{4}, y_{5}, \cdots) & = & W_{\pm}^{0}(y) + y_{4}W_{\pm}^{(4)}(y) + y_{5}W_{\pm}^{(5)}(y) + \cdots   \nonumber \\
       &  & + y_{4}^{2}W_{\pm}^{(4,4)}(y) + y_{4}y_{5}W_{\pm}^{(4,5)}(y) + \cdots,  \nonumber \\
       & = & \sum_{\boldkappa} W_{\pm}^{\boldkappa}(y) y^{[\boldkappa]},   \label{eq:2.15}
 \end{eqnarray}
in terms of the irrelevant scaled variables $y_{4}$, $y_{5}$, $\cdots$, where for brevity we have introduced the multi-index
 \begin{equation}
  \boldkappa = 0,\, (4),\, (5),\, \cdots,\, (4,4),\, (4,5),\, \cdots,\, (4,4,4),\, \cdots,   \label{eq:2.16}
 \end{equation}
and the associated conventions
 \begin{equation}
   y^{0} \equiv 1, \hspace{.3in} y^{[(i,j,\cdots,n)]} \equiv y_{i}y_{j}\cdots y_{n}.  \label{eq:2.17}
 \end{equation}
We also say $\boldkappa = [(i,j,\cdots,n)]$ is {\em odd} or {\em even} according as the sum $i + j + \cdots + n$ is odd or even. Then with an obvious extension of notation, the symmetry of $W_{\pm}(y, \cdots)$ requires
 \begin{equation}
  W_{\pm}^{\boldkappa}(-y) = (-)^{\boldkappa}W_{\pm}^{\boldkappa}(y).      \label{eq:2.18}
 \end{equation}

For small $y$ and $\tilde{t} > 0$ we can then write the further expansions
 \begin{eqnarray}
   W_{+}^{\boldkappa}(y) & = & W_{+0}^{\boldkappa} + y^{2}W_{+2}^{\boldkappa} + y^{4}W_{+4}^{\boldkappa} + \cdots, \hspace{.2in} \mbox{for $\boldkappa$ even},   \nonumber \\
       & = & yW_{+1}^{\boldkappa} + y^{3}W_{+3}^{\boldkappa} + y^{5}W_{+5}^{\boldkappa} + \cdots, \hspace{.2in} \mbox{for $\boldkappa$ odd}.     \label{eq:2.19}
 \end{eqnarray}
These series may, in general, be normalized via
 \begin{equation}
  W_{+2}^{0} = W_{+0}^{\boldkappa} = 1 \hspace{.2in}\mbox{($\boldkappa$ even)} \hspace{.2in}\mbox{or} \hspace{.2in} W_{+1}^{\boldkappa} = 1 \hspace{.2in} \mbox{($\boldkappa$ odd)},   \label{eq:2.20}
 \end{equation}
which serve to fix the nonuniversal metrical amplitudes $Q$, $U$ and $U_{k,e}$, etc. 

Note, however, that in setting $W_{+0}^{0} = W_{+2}^{0} = +1$ an appeal to thermodynamic convexity, \cite{ref4} together with $Q > 0$ and $\alpha > 0$, is entailed: see Ref. 17 where the consequences of the necessary convexity of the basic thermodynamic potentials is discussed both for the scaling functions and, more generally, for critical endpoints, thereby extending Schreinemakers' rules. \cite{sch,jcw}

For $\tilde{t} < 0$ the existence of the first-order transition leads to $|y|$ factors in the expansions so that one has
 \begin{equation}
  W_{-}^{\boldkappa}(y) = [W_{-0}^{\boldkappa} + |y|W_{-1}^{\boldkappa} + y^{2}W_{-2}^{\boldkappa} + |y|^{3}W_{-3}^{\boldkappa} + \cdots]\sigma_{\boldkappa}(y),  \label{eq:2.21}
 \end{equation}
where the special signum function is defined by
 \begin{eqnarray}
  \sigma_{\boldkappa}(y) & = & 1 \hspace{.2in}\mbox{for $\boldkappa$ even},  \nonumber \\
                     & = & \mbox{sgn}(y) \hspace{.2in}\mbox{for $\boldkappa$ odd}.      \label{eq:2.22}
 \end{eqnarray}
Convexity with $Q$, $U > 0$ then shows that $W_{-1}^{0}$ and $W_{-2}^{0}$ must both be {\em positive}: see Ref. 17.

For large arguments, $|y| \rightarrow \infty$, the individual scaling functions $W_{+}^{\boldkappa}(y)$ and $W_{-}^{\boldkappa}(y)$ must satisfy stringent matching conditions to ensure the analyticity of $G^{\beta\gamma}(g,T,h)$ through the surface $\tilde{t} = 0$ for all $h \neq 0$. These often overlooked conditions may be written
 \begin{equation}
  W_{\pm}^{\boldkappa}(y) \approx W_{\infty}^{\boldkappa}|y|^{(2-\alpha+\theta[\boldkappa])/\Delta}\left[1 + \sum_{l=1}^{\infty} w_{l}^{\boldkappa}(\pm|y|)^{-l/\Delta}\right] \sigma_{\boldkappa}(y),   \label{eq:2.23}
 \end{equation}
where the multiexponent $\theta[\boldkappa]$ is defined by
 \begin{equation}
  \theta[0] \equiv 0, \hspace{.2in} \theta[(i,j,\cdots,n)] = \theta_{i} + \theta_{j} + \cdots + \theta_{n},  \label{eq:2.24}
 \end{equation}
with $i$, $j$, $\cdots$, $n \geq 4$. By virtue of the normalizations (\ref{eq:2.20}) the numerical amplitudes $W_{-j}^{\boldkappa}$, $W_{+j}^{\boldkappa}$, $W_{\infty}^{\boldkappa}$, and $w_{l}^{\boldkappa}$ should all be universal (as should the exponents, $\alpha$, $\beta$, $\delta$, $\theta_{4}$, $\theta_{5}$, $\cdots$). Beyond that, as shown in Ref. 17, convexity dictates that $W_{\infty}^{0}$ and $w_{2}^{0}$ must be {\em positive} while $(w_{1}^{0})^{2}/w_{2}^{0}$ must be bounded above. The {\em sign} of $w_{1}^{0}$ is not determined by convexity alone but must, in general, be {\em negative}: see Ref. 17. This plays an important role in determining allowable density diagrams.

Finally, from (\ref{eq:2.13}) we note that the critical endpoint densities are
 \begin{equation}
  \rho_{1}^{\lambda e} = -G^{0}_{3} \hspace{.3in}\mbox{and}\hspace{.3in} \rho_{2}^{\lambda e} = -G^{0}_{1};   \label{eq:2.25}
 \end{equation}
see Fig. 3(b).

To close this section we recall from {\bf I} that the phase boundary $\sigma$ or $g = g_{\sigma}(T,h)$, follows by equating the two expressions $G = G^{\alpha}(g,T,h)$ and $G = G^{\beta\gamma}(g,T,h)$. Consequently, it is useful to define the thermodynamic potential difference
 \begin{equation}
  D(g,T,h) = G^{\alpha}(g,T,h) - G^{0}(g,T,h),   \label{eq:2.26}
 \end{equation}
which is noncritical by virtue of the definition of $G^{0}$ in (\ref{eq:2.13}). By our conventions the negative $g$ axis, i.e. $t = h = 0$, $g<0$, lies in the $\alpha$ phase (see Figs.~\ref{fig1} and~\ref{fig4}): this implies $D_{1} > 0$. The phase boundary $\rho$ and its extension $\tilde{\rho}$ above $T_{c}(g)$, is given by $\tilde{h}(g,T,h) = 0$. As in {\bf I}(5.4), we will assume that the $\lambda$-line is {\em not tangent} to the triple line $\tau$ at E. The densities $\rho_{1}$ and $\rho_{2}$ on the boundaries $\sigma$ and $\rho$ then follow from (\ref{eq:1.1}) and, by eliminating $g$ and $h$ at fixed $T$, the various isothermal binodals can be computed as expansions about $E$ or about $\lambda$: see Figs.~\ref{fig3} and~\ref{fig6}. We postpone the details until Sec.~\ref{sec4} and turn next to describing the results.

\section{The Endpoint Binodals and Their Inter-Relations}
\label{sec3}
In this section we describe the results of our analysis of the possible shapes of the various binodal curves and their inter-relations with one another as illustrated in Figs.~\ref{fig3} and~\ref{fig6}. After some preliminaries describing the ``rectification'' of the binodals, we consider first the behavior near the $\lambda$ line: this entails only the free energy $G^{\beta\gamma}(g,T,h)$ and, inasfar as the corrections to scaling are involved, extends previous knowledge somewhat. Then the binodals at the critical endpoint temperature $T = T_{e}$ are described: these are, perhaps, of most interest. The binodals associated with the $\sigma$ surface above $T_{e}$ are discussed next. These are analytic but their slopes and curvatures display critical singularities as $T\rightarrow T_{e}+$. Finally, the binodals associated with the three-phase triangle below $T_{e}$ are considered.

\subsection{Rectification of the Binodals}
\label{sec3.a}
We approach the description of the binodal curves by supposing that at fixed $T$ one may observe the densities $(\rho_{1},\rho_{2})$ of various pairs of coexisting phases. In {\em binary fluid} mixtures $\rho_{1}$ and $\rho_{2}$ might correspond directly to the number densities of the two species, B and C. In {\em ternary mixtures}, however, observations would normally be conducted at fixed temperature and pressure and  varying composition. Then $\rho_{1}$ and $\rho_{2}$ would each represent convenient {\em linear combinations} of the number densities of the three species, say, A, B, and C as represented typically in a triangle diagram. (Our analysis also applies to observations of {\em quaternary} mixtures {\em if sections} of the thermodynamic space corresponding to constant temperature, pressure, {\em and} a third field (or combination of chemical potentials) are constructed; however, experiments are not normally conducted that way and some further analysis would be needed to describe, say, a section at constant $T$, $p$, and $\rho_{3}$.)

We suppose next that the critical endpoint temperature $T_{e}$ itself can be determined with reasonable precision so that the variable $t = (T-T_{e})/T_{e}$ of (\ref{eq:2.1}) is well defined. Then the densities $(\rho_{1}^{\alpha e}, \rho_{2}^{\alpha e}) \equiv E^{\alpha}$ and $(\rho_{1}^{\lambda e}, \rho_{2}^{\lambda e}) \equiv E^{\lambda}$ of the spectator and critical phases {\em at} the endpoint can be found: see Figs.~\ref{fig3}(b) and~\ref{fig6}(b). These define an axis of slope
 \begin{equation}
  L_{\sigma} \equiv \Delta\rho_{1}/\Delta\rho_{2} = (\rho_{1}^{\lambda e} - \rho_{1}^{\alpha e})/(\rho_{2}^{\lambda e} - \rho_{2}^{\alpha e}).    \label{eq:3.1}
 \end{equation}
A natural second axis is found by noting that according to classical theory \cite{ref9} the critical binodals ${\cal B}_{e}^{\beta}$ and ${\cal B}_{e}^{\gamma}$ have a well defined common tangent at $E^{\lambda}$ of slope $(d\rho_{1}/d\rho_{2})^{e}_{{\cal B}_{e}^{\beta}} \equiv 1/L_{\rho}$, say. This is confirmed by our more general analyses which, indeed, predict that the binodals are flatter at $E^{\lambda}$ which eases the practical determination of $L_{\rho}$. (Note that it proves convenient to define $L_{\rho}$ reciprocally with respect to $L_{\sigma}$: see below. \cite{ref17})

To describe the various binodals near the endpoint it is then natural to adopt new density variables, $m$ and $\tilde{m}$, which are linearly related to $\rho_{1}$ and $\rho_{2}$ but utilize $E^{\lambda}$ as origin and are oriented along the axes just specified: see Figs.~\ref{fig3}(b) and~\ref{fig6}(b). Henceforth, therefore, we will utilize the {\em rectified density variables}
 \begin{eqnarray}
  m & = & \rho_{1} - \rho_{1}^{\lambda e} - L_{\sigma}(\rho_{2} - \rho_{2}^{\lambda e}),   \label{eq:3.2} \\
  \tilde{m} & = & \rho_{2} - \rho_{2}^{\lambda e} - L_{\rho}(\rho_{1} - \rho_{1}^{\lambda e}).  \label{eq:3.3}
 \end{eqnarray}
Furthermore, without loss of generality \cite{ref17} we assume that {\em the only pure phase located within the quadrant} $m > 0$, $\tilde{m} > 0$, {\em at} $T = T_{e}$ {\em is the} $\beta$ {\em phase}. Then, as illustrated in Figs.~\ref{fig3}(b) and~\ref{fig6}(b), the $\alpha$ phase at $T = T_{e}$ is restricted to $\tilde{m} < 0$ and only the $\gamma$ phase lies in the quadrant $m < 0$, $\tilde{m} > 0$.

The notation $m$ and $\tilde{m}$ is suggested by the magnetic case in which $m$, the magnetization, is the primary order parameter discontinuous across $\rho$ and coupling to the ordering field $h$, while $\tilde{m}$ is a secondary or subdominant order parameter conjugate to $g$. Note, indeed, that for {\em symmetric endpoints} we have \cite{ref17} $L_{\sigma} = L_{\rho} = 0$ so that if one shifts the definitions of the densities in a natural way to yield an origin $\rho_{1}^{\lambda e} = \rho_{2}^{\lambda e} = 0$ one simply has $m = \rho_{1}$ and $\tilde{m} = \rho_{2}$: see Fig.~\ref{fig6}.

\subsection{Lambda Line and Associated Binodals}
\label{sec3.b}
We note first (that within the postulates of Sec.~\ref{sec2}) the densities on the $\lambda$ line are {\em noncritical} functions of $T$ so that we have
 \begin{eqnarray}
  m_{\lambda}(T) & = & M_{1}t + M_{2}t^{2} + \cdots,   \label{eq:3.4} \\
  \tilde{m}_{\lambda}(T) & = & \tilde{M}_{1}t + \tilde{M}_{2}t^{2} + \cdots.  \label{eq:3.5}
 \end{eqnarray}
For a symmetric endpoint all the $M_{j}$ vanish identically. Beyond that, the coefficients $M_{j}$ and $\tilde{M}_{j}$ are not restricted in magnitude or sign although, of course, the $\lambda$ line itself cannot extend beyond the endpoint. Thus one must, here, have $t \leq 0$ in case {\bf A} and $t \geq 0$ in case {\bf B}.

Next notice that the binodals ${\cal B}_{<}^{\lambda \pm}$ for $T < T_{e}$, and --- see Fig.~\ref{fig6} --- ${\cal B}_{e}^{\lambda \pm}$ for $T = T_{e}$, and ${\cal B}_{>}^{\lambda \pm}$ for $T > T_{e}$, can all be treated together since by our postulates all of these binodals depend only on the free energy of the critical phase. Furthermore, inasfar as they are not truncated by the spectator-phase, they must all share the same singularities and vary uniformly with $T$. It is also convenient to describe the binodals with the aid of a parameter $s \geq 0$ (related to $|\tilde{t}|^{\beta}$) which vanishes on the $\lambda$ line and increases into the $\beta$ and $\gamma$ phases: coexisting phases correspond to the same value of $s$.

In the {\em symmetric case}, {\bf S}, the binodals associated with the $\lambda$ line or $\rho$ boundary may then be specified by
 \begin{eqnarray}
  m_{\pm} & = & \pm Bs\left[ 1 + b_{4}s^{\theta_{4}/\beta} + b_{1}s^{1/\beta} + b_{1,4}s^{(1+\theta_{4})/\beta} \right. \nonumber  \\
          &   & \hspace{.54in} \left. + \cdots + b_{5}s^{\delta + (\theta_{5}/\beta)} + \cdots \right],  \label{eq:3.6} \\
  \tilde{m} & = & \tilde{m}_{\lambda}(T) + \tilde{A}s^{(1-\alpha)/\beta}\left[ 1 + \tilde{a}_{4}s^{\theta_{4}/\beta} + \tilde{a}_{1}s^{1/\beta} + \cdots + \tilde{a}_{\bf n}s^{\tilde{\zeta}({\bf n})} + \cdots \right]  \nonumber \\
            &   & + \tilde{K}s^{1/\beta}\left[ 1 + k_{1}s^{1/\beta} + \cdots + k_{l}s^{l/\beta} + \cdots \right].    \label{eq:3.7}
 \end{eqnarray}
In (\ref{eq:3.6}) the general correction term has the form $b_{{\bf n}}(t)s^{\zeta ({\bf n})}$ where ${\bf n} = [n_{k}]$ is a multi-index with $n_{k} \geq 0$ and the exponents here and in (\ref{eq:3.7}) have the form
 \begin{eqnarray}
  \beta\tilde{\zeta}({\bf n}) & = & n_{0} + \sum_{j \geq 2} n_{2j}\theta_{2j},  \label{eq:3.8}  \\
  \beta\zeta ({\bf n}) & = & n_{0} + \sum_{j \geq 2}[n_{2j}\theta_{2j} + n_{2j + 1}(\Delta + \theta_{2j + 1})].  \label{eq:3.9} 
 \end{eqnarray}
The appearance of the exponent $\Delta = \beta\delta$ is due to the symmetry which acts to suppress the odd irrelevant variables.

The correction amplitudes $\tilde{a}_{4}(t)$, $\tilde{a}_{1}(t)$, $\cdots$, $b_{4}(t)$, $\cdots$ are noncritical but, generally, of indeterminate sign. However, the noncritical amplitude $B(t) = B_{e} + B_{2}t + \cdots$ is positive with our conventions and the signs $\pm$ correspond to the $\beta$ and $\gamma$ phases, respectively. The amplitudes $\tilde{A}(t) = \tilde{A}_{e} + \tilde{A}_{2}t + \cdots$ and $\tilde{K}(t) = \tilde{K}_{e} + \tilde{K}_{2}t + \cdots$ are also noncritical. For $\alpha > 0$, as we may assume here, the amplitude $\tilde{A}$ must be negative in case {\bf A} while it is positive in case {\bf B}. For $\alpha < 0$ the amplitude $\tilde{K}$ would have to have matching signs but that is not demanded for $\alpha > 0$. Explicit expressions for $\tilde{A}_{e}$, $B_{e}$, etc., are given in (\ref{eq:4.27}) and (\ref{eq:4.28}). 

It is clear by symmetry that the $(m,\tilde{m})$ {\em tielines} connecting coexisting phase points are all ``horizontal,'' that is, parallel to the $m$ axis $(\tilde{m} = 0)$: see Fig.~\ref{fig6}. Similarly, the {\em diameter} of the $\rho$ binodals, defined as the locus of midpoints of the tielines, is given simply by $m_{diam} = 0$, $\tilde{m}_{diam} \geq \tilde{m}_{\lambda}(T) \geq 0$.

The {\em symmetric} $\lambda$ {\em binodals} may finally be expressed directly in terms of $m$ as a variable by solving (\ref{eq:3.6}) for $s$ and substituting in (\ref{eq:3.7}). With $x = |m/B|$ this yields 
 \begin{eqnarray}
  \tilde{m} = \tilde{m}_{\lambda}(T) & + & \tilde{A}x^{(1-\alpha)/\beta}\left[ 1 + \bar{a}_{4}x^{\theta_{4}/\beta} + \bar{a}_{1}x^{1/\beta} + \cdots \right]   \nonumber \\
     & + & \tilde{K}x^{1/\beta}\left[ 1 + \bar{k}_{4}x^{\theta_{4}/\beta} + \bar{k}_{1}x^{1/\beta} + \cdots \right],   \label{eq:3.10}
 \end{eqnarray}
where $\bar{a}_{4} = \tilde{a}_{4} - (1-\alpha)b_{4}/\beta$ and so on. The term in $\tilde{A}$ provides the dominant behavior (when $\alpha > 0$) with $(1 - \alpha)/\beta \simeq 2.73$ for Ising $d = 3$ as quoted in the Introduction. However, the term in $\tilde{K}$ provides strongly competing corrections of relative order $|m|^{\alpha/\beta}$: note that $1/\beta \simeq 3.05$. The higher order correction terms run through all powers of $x$ with exponents of the form $\zeta ({\bf n}_{1}) + \zeta ({\bf n}_{2}) + \cdots + \zeta ({\bf n}_{l})$.

The {\em nonsymmetric}, {\bf N}, binodals associated with $\lambda$ line or $\rho$ surface --- see Fig.~\ref{fig3}(a) --- may be described similarly. In terms of the parameter $s$ we find
 \begin{eqnarray}
  m_{\pm} & = & m_{\lambda}(T) \pm Bs\left[ 1 + b_{4}s^{\theta_{4}/\beta} + b_{1}s^{1/\beta} + \cdots \pm b_{5}s^{\theta_{5}/\beta} \pm \cdots \right]   \nonumber \\
          &   & + As^{(1-\alpha)/\beta}\left[ 1 + a_{4}s^{\theta_{4}/\beta} + a_{1}s^{1/\beta} + \cdots \pm a_{5}s^{\theta_{5}/\beta} + \cdots \right] \nonumber \\
          &   & + Ks^{1/\beta}\left[ 1 + \cdots + k_{l}s^{l/\beta} + \cdots \right],   \label{eq:3.11} \\
  \tilde{m}_{\pm} & = & \tilde{m}_{\lambda}(T) + \tilde{A}s^{(1-\alpha)/\beta}\left[ 1 + \tilde{a}_{4}s^{\theta_{4}/\beta} + \tilde{a}_{1}s^{1/\beta} + \cdots \pm \tilde{a}_{5}s^{\theta_{5}/\beta} \pm \cdots \right]   \nonumber \\
                  &   & + \tilde{K}s^{1/\beta}\left[ 1 + \cdots + \tilde{k}_{l}s^{l/\beta} + \cdots \right] \pm \tilde{B}s^{(1+\beta)/\beta}\left[ 1 + \tilde{b}_{4}s^{\theta_{4}/\beta} + \cdots \pm \tilde{b}_{5}s^{\theta_{5}/\beta} \pm \cdots \right]   \nonumber \\
                  &   & \pm B^{\prime} ts\left[ 1 + b_{4}^{\prime} s^{\theta_{4}/\beta} + \cdots \pm b_{5}^{\prime} s^{\theta_{5}/\beta} \pm \cdots \right],       \label{eq:3.12}
 \end{eqnarray}
where, again, all the coefficients are noncritical and the same remarks as before apply to the signs of $\tilde{A}$, $\tilde{K}$, and $B$. The correction factors for the $A$, $\tilde{A}$, $B$, $\tilde{B}$ and $B^{\prime}$ terms run through all powers of $s$ with exponents of the form $(n_{0} + \theta[{\boldkappa}])/\beta$ (recalling the definitions (\ref{eq:2.16}), (\ref{eq:2.24}), etc.); terms with odd $\boldkappa$ carry $\pm$ signs; when $n_{0} = 0$ we have $\tilde{a}_{\boldkappa} = a_{\boldkappa}$, and $\tilde{b}_{\boldkappa} = b_{\boldkappa}$. Expressions for $A$, $\tilde{A}$, etc. are given in (\ref{eq:4.29})-(\ref{eq:4.32}).

Now note that the amplitude $B^{\prime}$ carries a factor $t$ which vanishes at $T_{e}$. Away from the endpoint this term induces a linear variation of $\tilde{m}$ with $m$ which simply means that the tangents to the binodals at the $\lambda$ point (for $T \neq T_{e}$) are no longer parallel to the tangent at the endpoint. Such a variation is, of course, to be expected and does not represent any real change of shape as $T$ deviates from $T_{e}$. To see this more explicitly, note that we may redefine the coefficient $L_{\rho}$, which enters the definition (\ref{eq:3.3}) of $\tilde{m}$, as a noncritical function, $L_{\rho}(t)$, chosen so that the tangent at the $\lambda$ point is always parallel to $\tilde{m} = 0$ (i.e., to the $m$ axis); then one has $B^{\prime} \equiv 0$ while the other terms in (\ref{eq:3.12}) do not change form. With this understanding for $t \neq 0$ we may conveniently define
 \begin{equation}
  \Delta m = m - m_{\lambda}(t) \hspace{.2in} \mbox{and} \hspace{.2in} \Delta\tilde{m} = \tilde{m} - \tilde{m}_{\lambda}(t),  \label{eq:3.13}
 \end{equation}
which reduce to $m$ and $\tilde{m}$, respectively, at the endpoint.

The {\em diameters} of the nonsymmetric $\lambda$ binodals may now be found parametrically by multiplying out in (\ref{eq:3.11}) and (\ref{eq:3.12}) and dropping all terms which carry $\pm$ signs. If the parameter $s$ is eliminated in favor of $\tilde{x} = \Delta\tilde{m}_{diam}/\tilde{A}$, the diameters can be written
 \begin{eqnarray}
  \Delta m_{diam} & = & A\tilde{x}\left[ 1 + {\cal K}_{\lambda}\tilde{x}^{\alpha/(1-\alpha)} + \cdots + a_{4}\tilde{x}^{\theta_{4}/(1-\alpha)} + \cdots \right]  \nonumber \\
                  &   & + {\cal U}_{\lambda}\tilde{x}^{(\beta + \theta_{5})/(1-\alpha)}\left[1 + \cdots \right],   \label{eq:3.14}
 \end{eqnarray}
where we suppose $\alpha > 0$ while
 \begin{equation}
  {\cal K}_{\lambda} = (\tilde{A}K - A\tilde{K})/\tilde{A}A \hspace{.2in} \mbox{and} \hspace{.2in} {\cal U}_{\lambda} = Bb_{5}.   \label{eq:3.15}
 \end{equation}
We see that the slope $(\partial m/\partial \tilde{m})$ of the diameter remains finite at the endpoint but, in general, the curvature {\em diverges} at the endpoint.

The slopes $\Sigma_{\lambda} = \Delta\tilde{m}/\Delta m$ of the tielines follow similarly from the terms in (\ref{eq:3.11}) and (\ref{eq:3.12}) carrying the $\pm$ signs. Using, again, $\tilde{x} = \Delta\tilde{m}_{diam}/\tilde{A}$ as variable one finds, for $\alpha > 0$, 
 \begin{eqnarray}
  \Sigma_{\lambda} & = & \frac{\tilde{B}}{B}\tilde{x}^{1/(1-\alpha)} \left[ 1 - \frac{\tilde{K}}{(1-\alpha)\tilde{A}}\tilde{x}^{\alpha/(1-\alpha)} + (\tilde{b}_{4}-b_{4})\tilde{x}^{\theta_{4}/(1-\alpha)} + \cdots \right.  \nonumber \\
   &  & \hspace{1.in} + \left. \frac{\tilde{A}}{\tilde{B}}\tilde{x}^{(\theta_{5}+\Delta -2)/(1-\alpha)} + \cdots \right].  \label{eq:3.16}
 \end{eqnarray}
As was anticipated, the tielines do not, in general, remain parallel to the $\lambda$-point binodal tangent; however, the variation in slope is evidently {\em slower} than linear in $\Delta\tilde{m}$.

Finally, one may eliminate $s$ between (\ref{eq:3.11}) and (\ref{eq:3.12}) directly and write the general, {\em nonsymmetric} $\lambda$ {\em binodals} in terms of $x = |\Delta m/ B|$ as
 \begin{eqnarray}
  \Delta\tilde{m} & = & \tilde{A}x^{(1-\alpha)/\beta}\left[ 1 \pm a_{B}\tilde{B}x^{(\alpha+\beta)/\beta} + \bar{a}_{4}x^{\theta_{4}/\beta} \pm a_{A}Ax^{(\Delta - 1)/\beta} \right.  \nonumber \\
                  &   & \hspace{.9in} \left. \pm a_{K}Kx^{(1-\beta)/\beta} + \cdots \pm \bar{a}_{5}x^{\theta_{5}/\beta} + \cdots \right]  \nonumber  \\
                  &   & + \tilde{K}x^{1/\beta}\left[ 1 + \bar{b}_{4}x^{\theta_{4}/\beta} \pm \bar{b}_{A}Ax^{(\Delta -1)/\beta} \pm \bar{b}_{K}Kx^{(1-\beta)/\beta} + \cdots \right],  \label{eq:3.17}
 \end{eqnarray}
where the $\pm$ signs refer to $\Delta m \gtrless 0$ (for $B > 0$) while
 \begin{eqnarray}
   a_{A} & = & a_{K} = -(1-\alpha)/\beta B, \hspace{.3in} a_{B} = 1/\tilde{A}, \nonumber \\
   \bar{a}_{4} & = & \tilde{a}_{4} - (1-\alpha)b_{4}/\beta, \hspace{.3in} \bar{b}_{4} = -b_{4}/\beta, \cdots.  \label{eq:3.18}
 \end{eqnarray}

We see that the leading behavior of the binodals, with exponent $(1-\alpha)/\beta$ (for $\alpha > 0$), is the same as in the symmetric case (\ref{eq:3.10}). The surprising new feature, however, is the large number of numerically similar low-order correction terms. If we write the expansion for a general binodal in the form
 \begin{equation}
   \Delta\tilde{m} = \sum_{i} {\cal A}_{i}^{\pm}|m|^{\psi_{i}},   \label{eq:3.19}
 \end{equation}
(with $\pm$ for $m \gtrless 0$) the nonsymmetric $\lambda$-line binodals generate the exponent sequence
 \begin{eqnarray}
  \psi_{i}^{\lambda}\beta & = & 1-\alpha,\hspace{.2cm} 1,\hspace{.2cm} 1+\beta,\hspace{.2cm} 1-\alpha + \theta_{4},\hspace{.2cm} 2 - 2\alpha - \beta,\hspace{.2cm} 1 + \theta_{4},\hspace{.2cm} 2 - \alpha - \beta,  \nonumber \\
                          &   & 2 - \beta,\hspace{.2cm} \cdots,\hspace{.2cm} 1 - \alpha + \theta_{5},\hspace{.2cm} \cdots.  \label{eq:3.20}
 \end{eqnarray}
For $d=3$ the Ising numerical values are
 \begin{eqnarray}
  \psi_{i}^{\lambda}\beta & \simeq & 0.89_{5},\, 1,\, 1.32_{8},\, 1.44,\, 1.46,\, 1.54,\, 1.57,  \nonumber \\
                          &  &  1.67,\, \cdots,\, 1.9,\, \cdots,   \label{eq:3.21}
 \end{eqnarray}
where, here and below, we use the {\em rough} approximation $\theta_{5} \simeq 1.0$; when $d \rightarrow 4-$ one gets 1, 1, $\frac{3}{2}$, 1, $\frac{3}{2}$, 1, $\frac{3}{2}$, $\frac{3}{2}$, $\cdots$, $\frac{3}{2}$, $\cdots$. Finally, note that the presence of the various $\pm$ signs in (\ref{eq:3.17}) reflects the nontrivial behavior of the diameter and consequent lack of binodal symmetry outside the innermost asymptotic region.

\subsection{Spectator Phase Boundary at the Endpoint}
\label{sec3.c}
The spectator phase, $\alpha$, is bounded in the space of thermodynamic fields by the surface $\sigma$ (see Figs.~\ref{fig1} and~\ref{fig4}) which may be specified by the function $g_{\sigma}(t,h)$ which, as explained in {\bf I}, is found by equating the $\alpha$ and $\beta\gamma$ free energies, i.e., by solving $G^{\alpha}(g_{\sigma},t,h) = G^{\beta\gamma}(g_{\sigma},t,h)$. In leading order this was carried out in {\bf I} but, for the present purposes it is useful to have the results correct to higher order. Here we present expressions for $T = T_{e}$ (or $t = 0$), i.e. on the endpoint isotherm.

The detailed analysis is presented in Sec.~\ref{sec4}.C where one sees that it is advantageous to retain $\tilde{h}$ as a principal variable. The results for the {\em symmetric case} are the simplest in form: we find
 \begin{equation}
  g_{\sigma}(t=0;\tilde{h}) = -J|\tilde{h}|^{(\delta +1)/\delta}Z_{S}(|\tilde{h}|) - J_{2}\tilde{h}^{2} - J_{4}|\tilde{h}|^{2+(2/\delta)} + \cdots,  \label{eq:3.22}
 \end{equation}
where the singular correction factor is
 \begin{eqnarray}
  Z_{S}(z) = 1 & \pm & c_{1}z^{(1-\alpha)/\Delta} + c_{2}z^{2(1-\alpha)/\Delta} + c_{3}z^{2-(1/\Delta)} \nonumber \\
           & + & c_{4}z^{\theta_{4}/\Delta} \pm c_{4}^{\prime} z^{(1-\alpha+\theta_{4})/\Delta} + c_{5}z^{1 + (\theta_{5}/\Delta)} + \cdots.   \label{eq:3.23}
 \end{eqnarray}
The {\em upper} (plus) signs in $Z_{S}$ correspond to case {\bf B} or $q_{1} < 0$; recall (\ref{eq:1.4}) and Fig.~\ref{fig4}; the {\em lower} (minus) signs describe case {\bf A} when $q_{1} > 0$: see (\ref{eq:1.3}) and Fig.~\ref{fig1}.

The leading amplitude in (\ref{eq:3.22}) is given, using (\ref{eq:2.26}), by
 \begin{equation}
  J = Q_{e}U^{(\delta + 1)/\delta}W_{\infty}^{0}/(D_{1} - r_{0}D_{3}),  \label{eq:3.24}
 \end{equation}
where $Q_{e}$ and $U$ are defined via (\ref{eq:2.13}) and (\ref{eq:2.10}) while, for the symmetric case, one has $r_{0}D_{3} = 0$ and $J > 0$. In addition we state
 \begin{equation}
  c_{1} = w_{1}^{0}|q_{1}|J/U^{1/\Delta}, \hspace{.3in} c_{2} = w_{2}^{0}q_{1}^{2}J^{2}/U^{2/\Delta}, \label{eq:3.25}
 \end{equation}
while the other coefficients are recorded in Sec.~\ref{sec4}.C. The result (\ref{eq:3.22}) can be expressed in terms of $h$ by using
 \begin{equation}
  \tilde{h} = h[ 1 - r_{1}J|h|^{(2-\alpha)/\Delta} \mp r_{1}c_{1}J|h|^{(3-2\alpha)/\Delta} + \cdots ] \label{eq:3.26}
 \end{equation}
which, however, is valid {\em only} for $t = 0$ and $g = g_{\sigma}$. We note that $(\delta + 1)/\delta = (2 - \alpha)/\Delta \simeq 1.21$ in agreement with {\bf I}: see also Fig.~\ref{fig5} for a portrayal of $g_{\sigma}(0,h)$. We defer discussion of the correction exponents until the noncritical/spectator binodals are presented: see Eqs. (\ref{eq:3.36}) and (\ref{eq:3.44}).

In the {\em nonsymmetric case} the leading variation of $g_{\sigma}(t=0)$ is, in general, {\em linear} in $\tilde{h}$ (and $h$): see Fig.~\ref{fig2}. Specifically, subject to
 \begin{equation}
  J_{1} \equiv D_{3}/(D_{1}-r_{0}D_{3}) \neq 0,\hspace{.1in} \infty,  \label{eq:3.27}
 \end{equation}
we find
 \begin{eqnarray}
  g_{\sigma}(t=0;\tilde{h}) & = & -J_{1}\tilde{h} - J|\tilde{h}|^{(\delta +1)/\delta}Z_{N}(|\tilde{h}|) - J_{2}\tilde{h}^{2}  \nonumber \\
     &  & - J_{3}\tilde{h}|\tilde{h}|^{(\delta +1)/\delta} - J_{4}|\tilde{h}|^{2(\delta +1)/\delta} + \cdots,   \label{eq:3.28}
 \end{eqnarray}
where the {\em non}symmetric singular factor has the expansion
 \begin{eqnarray}
  Z_{N}(z) & = & 1 + \tilde{\sigma}_{t}d_{1}z^{1-(1/\Delta)} - d_{1}^{\prime} z^{(1-\alpha)/\Delta} + d_{2}z^{2-(2/\Delta)} - (\tilde{\sigma}_{t}d_{1}d_{1}^{\prime} + \tilde{\sigma}_{h}d_{2}^{\prime})z^{(\Delta-\alpha)/\Delta}  \nonumber \\
           &   & + (d_{2}^{\prime\prime} + d_{1}^{\prime 2})z^{2(1-\alpha)/\Delta} + \tilde{\sigma}_{t}d_{3}z^{3-(3/\Delta)} - (d_{3}^{\prime} + \tilde{\sigma}_{t}d_{3}^{\prime\prime})z^{2-(1+\alpha)/\Delta} + d_{3}^{\prime\prime\prime} z^{2-(1/\Delta)}  \nonumber \\
           &   & + d_{4}z^{\theta_{4}/\Delta} + \tilde{\sigma}_{t}d_{4}^{\prime} z^{(\theta_{4}+\Delta-1)/\Delta} - (d_{4}^{\prime\prime} + d_{1}^{\prime}d_{4})z^{(\theta_{4}+1-\alpha)/\Delta} + \tilde{\sigma}_{h}d_{5}z^{\theta_{5}/\Delta}  \nonumber \\
           &   & + \tilde{\sigma}_{t}\tilde{\sigma}_{h}d_{5}^{\prime} z^{(\theta_{5}+\Delta-1)/\Delta} - \tilde{\sigma}_{h}(d_{5}^{\prime\prime} + d_{1}^{\prime} d_{5})z^{(\theta_{5}+1-\alpha)/\Delta} - \cdots.  \label{eq:3.29}
 \end{eqnarray}
The two signum factors are given by
 \begin{equation}
  \tilde{\sigma}_{t} = \mbox{sgn}(\tilde{t}) = \mbox{sgn}(\tilde{q}h), \hspace{.3in} \tilde{\sigma}_{h} = \mbox{sgn}(\tilde{h}) = \mbox{sgn}(j_{1}h),  \label{eq:3.30}
 \end{equation}
in which we suppose the coefficients
 \begin{equation}
  \tilde{q} = q_{0} - q_{1}(D_{3}/D_{1}), \hspace{.3in} j_{1} = 1 - r_{0}(D_{3}/D_{1}), \label{eq:3.31}
 \end{equation}
are nonvanishing; this will be true in the general nonsymmetric case. (We do not analyze the exceptions although no problems of principle arise.)

We see from (\ref{eq:3.28})-(\ref{eq:3.30}) that terms which change sign are not now determined simply by the slope of the $\lambda$ line (case {\bf A} or case {\bf B}), as in the symmetric situation, but rather by more complicated considerations. This arises simply because the manifold $\tilde{t} = 0$ in the $(g,t,h)$ space, see Fig.~\ref{fig1}, can cut the plane $t=0$ in various ways. For small asymmetry, $j_{1}$ remains positive giving $\tilde{\sigma}_{h} = \mbox{sgn}(h)$ but $\tilde{q}$ may be of either sign. As expected from {\bf I}, the leading singularity in $g_{\sigma}$ is the same as in the symmetric situation; however, the corrections now contain further, new powers.

The leading correction amplitudes in $Z_{N}$ are
 \begin{equation}
  d_{1} = w_{1}^{0}|\tilde{q}|/|j_{1}|U^{1/\Delta}, \hspace{.3in} d_{1}^{\prime} = w_{1}^{0}(q_{1}-q_{0}r_{0})J/U^{1/\Delta}.   \label{eq:3.32}
 \end{equation}
The remaining leading coefficients are listed in Sec.~\ref{sec4}.C. As before the result (\ref{eq:3.28}) can be expressed in terms of $h$ by making the substitution
 \begin{equation}
  \tilde{h} = j_{1}h - j|h|^{(\delta +1)/\delta} + j^{\prime} h|h|^{2/\delta} - \tilde{\sigma}_{t}j^{\prime\prime} |h|^{(3-2\alpha-\beta)/\Delta} - j_{2}h^{2} + \cdots,   \label{eq:3.33}
 \end{equation}
where $j = r_{0}Jj_{1}|j_{1}|^{(\delta +1)/\delta}$ while $j^{\prime}$, etc., are given below in (\ref{eq:4.49}).

\subsection{Noncritical Endpoint Binodals}
\label{sec3.d}
We are now in a position to answer Widom's question regarding the shape of the noncritical or spectator-phase binodals, ${\cal B}_{e}^{\alpha\pm}$, {\em at} the endpoint. The essential point is that the densities $\rho_{1}$ and $\rho_{2}$ and, hence, $m$ and $\tilde{m}$, are {\em noncritical} functions of $g$, $t$, and $h$ in the spectator-phase $\alpha$ since $G^{\alpha}(g,t,h)$ is noncritical. Consequently, on the endpoint isotherm, $t=0$, the singular shape of the $\alpha$ binodals directly reflects the singular shape of the phase boundary $g_{\sigma}(0,h)$.

To state the results for the {\em symmetric case} we introduce the endpoint susceptibilities
 \begin{equation}
  \chi_{e}^{\alpha} = -2G^{\alpha}_{9} > 0 \hspace{.2in} \mbox{and} \hspace{.2in} \tilde{\chi}_{e}^{\alpha} = -2G^{\alpha}_{4} > 0,   \label{eq:3.34}
 \end{equation}
and the endpoint density
 \begin{equation}
  \tilde{m}^{\alpha}_{e} = -G^{\alpha}_{1} + G^{0}_{1} < 0.  \label{eq:3.35}
 \end{equation}
The noncritical binodal is then given by
 \begin{equation}
  \tilde{m} = \tilde{m}^{\alpha}_{e} - C|x_{\alpha}|^{(\delta+1)/\delta}Z_{S}(|x_{\alpha}|) - C_{2}x_{\alpha}^{2} - C_{3}x_{\alpha}^{2(\delta+1)/\delta} + \cdots,  \label{eq:3.36}
 \end{equation}
where
 \begin{equation}
  x_{\alpha} = m/\chi_{e}^{\alpha}, \hspace{.2in} C = J\tilde{\chi}_{e}^{\alpha}, \hspace{.2in} C_{2} = D_{9}\tilde{\chi}_{e}^{\alpha}/D_{1}, \hspace{.2in} C_{3} = \left(\frac{D_{4}}{D_{1}} - \frac{Q_{1}}{Q_{e}}\right)J^{2}\tilde{\chi}_{e}^{\alpha},  \label{eq:3.37}
 \end{equation}
while $Z_{S}(z)$ is given in (\ref{eq:3.23}). The leading exponent is $(\delta+1)/\delta = (2-\alpha)/\Delta$ as stated in the Introduction. If we use the general binodal expansion (\ref{eq:3.19}) the sequence of exponents arising now is
 \begin{eqnarray}
  \psi_{i}^{\alpha}\Delta & = & 2-\alpha,\hspace{.2cm} 2-\alpha+\theta_{4},\hspace{.2cm} 3-2\alpha,\hspace{.2cm} 4-2\alpha-2\beta,\hspace{.2cm} 3-2\alpha+\theta_{4},\hspace{.2cm} 4-3\alpha,  \nonumber \\
   &   & 5-3\alpha-2\beta,\hspace{.2cm} \cdots,\hspace{.2cm} 4-2\alpha-\beta+\theta_{5},\hspace{.2cm} \cdots,  \hspace{.2in} \mbox{({\bf S})}, \label{eq:3.38}
 \end{eqnarray}
with values for Ising $d=3$:
 \begin{equation}
  \psi_{i}^{\alpha} \simeq 1.20_{9},\, 1.55,\, 1.78,\, 2,\, 2.13,\, 2.35,\, 2.57,\, \cdots,\, 2.9,\, \cdots,\hspace{.2in} \mbox{({\bf S})} \label{eq:3.39}
 \end{equation}
(using, again, $\theta_{5} \simeq 1.0$).

In the limit $d\rightarrow 4-$ the sequence for $\psi_{i}^{\alpha}$ is $\frac{4}{3}$, $\frac{4}{3}$, 2, 2, 2, $\frac{8}{3}$, $\frac{8}{3}$, $\cdots$, $\frac{8}{3}$, $\cdots$. Note that the leading correction exponent found by Klinger \cite{ref9} was $\frac{5}{3}$. His classical phenomenological treatment should correspond to $d \rightarrow 4-$ but $\frac{5}{3}$ does {\em not} appear here: the reason is that he did not (expressly) consider the symmetric situation. We also find the exponent $\frac{5}{3}$ (and others) when symmetry is lacking.

In the {\em nonsymmetric} case, the endpoint susceptibilities become more complicated: we find they are given by
 \begin{eqnarray}
  \chi_{e}^{\alpha} & = & -2(G^{\alpha}_{9} - 2L_{\sigma}G^{\alpha}_{5} + L_{\sigma}^{2}G^{\alpha}_{4}) > 0,  \label{eq:3.40}  \\
  \tilde{\chi}_{e}^{\alpha} & = & -2(G^{\alpha}_{4} - 2r_{0}G^{\alpha}_{5} + r_{0}^{2}G^{\alpha}_{9}) > 0, \label{eq:3.41}
 \end{eqnarray}
where the significance of the axis slope, $L_{\sigma}$, was explained in Sec.~\ref{sec3}.A above. From (\ref{eq:3.1}), (\ref{eq:2.4}) and (\ref{eq:2.25}) we obtain
 \begin{equation}
  L_{\sigma} = (G^{\alpha}_{3} - G^{0}_{3})/(G^{\alpha}_{1}-G^{0}_{1}),   \label{eq:3.42}
 \end{equation}
while the endpoint density is
 \begin{equation}
  \tilde{m}^{\alpha}_{e} = r_{0}(G^{\alpha}_{3}-G^{0}_{3}) - G^{\alpha}_{1} + G^{0}_{1} < 0.  \label{eq:3.43}
 \end{equation}

Using, again, $x_{\alpha} = m^{\alpha}/\chi^{\alpha}_{e}$ as a variable, the noncritical endpoint binodal
in the {\em nonsymmetric} case is expressed by
 \begin{equation}
  \tilde{m}^{\alpha} = \tilde{m}^{\alpha}_{e} + \tilde{m}^{\alpha}_{1}x_{\alpha} + \tilde{m}^{\alpha}_{2}|x_{\alpha}|^{(\delta +1)/\delta} \pm \tilde{m}^{\alpha}_{3}|x_{\alpha}|^{(\delta +2)/\delta} + \cdots, \label{eq:3.44}
 \end{equation}
where $\pm$ corresponds to $h \gtrless 0$ while the amplitudes $\tilde{m}^{\alpha}_{1}$, $\tilde{m}^{\alpha}_{2}$, $\cdots$ are presented below. The linear variation of $\tilde{m}^{\alpha}$ with $x_{\alpha}$ shows that the tangent to the noncritical endpoint binodal at the endpoint $E^{\alpha}$ is, in general, {\em not} parallel to the tangent at $E^{\lambda}$ (the $m$-axis): see Fig.~\ref{fig3}(b). The corresponding amplitude, $\tilde{m}^{\alpha}_{1}$, is 
 \begin{equation}
  \tilde{m}^{\alpha}_{1} = 2(-G^{\alpha}_{5} + r_{0}G^{\alpha}_{9}) + 2L_{\sigma}(G^{\alpha}_{4} - r_{0}G^{\alpha}_{5}).  \label{eq:3.45}
 \end{equation}
The leading singular exponent, namely $1+(1/\delta)$, is evidently the same as the symmetric case, while the amplitude, $\tilde{m}^{\alpha}_{2}$, is 
 \begin{equation}
  \tilde{m}^{\alpha}_{2} = 2(jJ_{1} - J)[ -G^{\alpha}_{4} + r_{0}G^{\alpha}_{5} - \tilde{m}^{\alpha}_{1}(-G^{\alpha}_{5} + L_{\sigma}G^{\alpha}_{4})/\chi^{\alpha}_{e}].   \label{eq:3.46}
 \end{equation}
Recall that the coefficients $j$, $J_{1}$, and $J$ are defined above in Sec.~\ref{sec3}.C.

The leading correction exponent is now just that found by Klinger \cite{ref9} in his classical treatment; it does {\em not} appear in the symmetric case. The expression for its amplitude, $\tilde{m}^{\alpha}_{3}$, is complicated but, for the record, we quote the result, namely,
 \begin{eqnarray}
  \tilde{m}^{\alpha}_{3} & = & \frac{(\delta +1)}{\delta}\left(\frac{\chi^{\alpha}_{1}}{\chi^{\alpha}_{e}}\right)\left[ \tilde{m}^{\alpha}_{1}\frac{\chi^{\alpha}_{1}}{\chi^{\alpha}_{e}} - 2(jJ_{1} - J)(-G^{\alpha}_{4} + r_{0}G^{\alpha}_{5})\right]  \nonumber \\
   &   & + 2\mbox{sgn}(j_{1})g_{2}\left[ (-G^{\alpha}_{5} + L_{\sigma}G^{\alpha}_{4})\frac{\tilde{m}^{\alpha}_{1}}{\chi^{\alpha}_{e}} - (-G^{\alpha}_{4} + r_{0}G^{\alpha}_{5})\right],  \label{eq:3.47}
 \end{eqnarray}
where the new coefficients are
 \begin{eqnarray}
  \chi^{\alpha}_{1} & = & 2(-G^{\alpha}_{5} + L_{\sigma}G^{\alpha}_{4})(jJ_{1} - J),   \label{eq:3.48} \\
  g_{2} & = & r_{0}^{2}|j_{1}|^{3+(2/\delta)} J^{2}J_{1} - \frac{(\delta + 1)}{\delta}|j_{1}|^{1/\delta}jJ.  \label{eq:3.49}
 \end{eqnarray} 

\subsection{Critical Endpoint Binodals}
\label{sec3.e}
Now we conclude our discussion of the endpoint itself by presenting, finally, the shape of the critical phase binodals, ${\cal B}_{e}^{\beta}$ and ${\cal B}_{e}^{\gamma}$. These can be obtained by using the thermodynamic potential for the critical phase, $G^{\beta\gamma}(g,t,h)$, and the endpoint phase boundary, $g_{\sigma}(\tilde{h})$. Details are given in Sec.~\ref{sec4}.D. As discussed before, it is convenient to describe the binodals with the aid of a parameter $s$ (in this case, related to $|\tilde{h}|^{\beta/\Delta}$) which vanishes at the endpoint and increases into the $\beta$ and $\gamma$ phases.

In the {\em symmetric} case, {\bf S}, the critical endpoint binodals may then be specified by
 \begin{eqnarray}
  m & = & \pm E s[1+u_{4}s^{\theta_{4}/\beta} + u_{1}s^{(1-\alpha)/\beta} + \cdots] \pm V_{1}s^{\Delta/\beta}[1+v_{1}s^{\Delta/\beta}+\cdots]  \nonumber  \\
    &   & \pm V_{2}s^{(1-\alpha +\Delta)/\beta}[1+\cdots] \pm V_{3}s^{(2-\alpha + \Delta)/\beta}[1+\cdots],  \label{eq:3.50}  \\
\vspace{.7cm}
  \tilde{m} & = & \tilde{E}s^{(1-\alpha)/\beta}[1+\tilde{u}_{4}s^{\theta_{4}/\beta} + \tilde{u}_{1}s^{(1-\alpha)/\beta} + \cdots] + \tilde{V}s^{(2-\alpha)/\beta}[1+\cdots],   \label{eq:3.51}
 \end{eqnarray}
where $\pm$ corresponds to $\tilde{h}\gtrless 0$, and the coefficients are given in (\ref{eq:4.59}) and (\ref{eq:4.60}).

The {\em symmetric} critical endpoint binodals may finally be expressed in terms of $m$ as a variable by solving (\ref{eq:3.50}) for $s$ and substituting in (\ref{eq:3.51}). With $x=|m/E|$ this yields
 \begin{equation}
  \tilde{m} = \tilde{E}x^{(1-\alpha)/\beta}[1+\bar{u}_{4}x^{\theta_{4}/\beta} + \bar{u}_{1}x^{(1-\alpha)/\beta} + \cdots], \label{eq:3.52}
 \end{equation}
where
 \begin{equation}
   \bar{u}_{1} = \tilde{u}_{1} - (1-\alpha)u_{1}/\beta, \hspace{.3in} \bar{u}_{4} = \tilde{u}_{4} - (1-\alpha)u_{4}/\beta.  \label{eq:3.53}
 \end{equation}
The term in $\tilde{E}$ provides the dominant behavior with the same exponent as the $\lambda$-line binodals given in Sec.~\ref{sec3}.B. One should notice that the amplitude $\tilde{E}$ is negative in case {\bf A} while it is positive in case {\bf B} owing to the negative sign of $w_{1}^{0}$ discussed following (\ref{eq:2.23}).\cite{mef:yck} Hence it has the same sign as the amplitude $\tilde{A}$ of the lambda line binodals: see (\ref{eq:3.10}). This also holds in the nonsymmetric case.

Indeed, the {\em nonsymmetric}, {\bf N}, critical endpoint binodals may be described similarly. In terms of the parameter $s$ we find
 \begin{eqnarray}
  m & = & \pm Es[1 \pm u_{1}s^{(\Delta-1)/\beta} + u_{2}s^{(1-\alpha)/\beta} + \cdots + u_{4}s^{\theta_{4}/\beta} + \cdots \pm u_{5}s^{\theta_{5}/\beta} + \cdots]  \nonumber \\
    &   &  + V_{1}s^{(1-\alpha)/\beta}[1 \pm v_{1}s^{(\Delta-1)/\beta} + v_{2}s^{(1-\alpha)/\beta} + \cdots  + v_{4}s^{\theta_{4}/\beta} + \cdots \pm v_{5}s^{\theta_{5}/\beta} + \cdots]  \nonumber \\
    &   & \pm V_{2}s^{\Delta/\beta}[1 \pm v_{0}s \cdots] + V_{3}s^{(2-\alpha)/\beta}[1+\cdots],  \label{eq:3.54} \\
  \tilde{m} & = & \tilde{E}s^{(1-\alpha)/\beta}[1 \pm \tilde{u}_{1}s^{(\Delta -1)/\beta} + \tilde{u}_{2}s^{(1-\alpha)/\beta} + \cdots + \tilde{u}_{4}s^{\theta_{4}/\beta} \cdots \pm \tilde{u}_{5}s^{\theta_{5}/\beta} + \cdots]  \nonumber \\
  &   & \pm \tilde{V}_{1}s^{\Delta/\beta}[1 \pm \tilde{v}_{0}s + \cdots] + \tilde{V}_{2}s^{(2-\alpha)/\beta}[1 + \cdots],  \label{eq:3.55}
 \end{eqnarray}
where the leading coefficients are presented in (\ref{eq:4.61})-(\ref{eq:4.63}). Solving for $s$ in (\ref{eq:3.54}) and substituting into (\ref{eq:3.55}), one finally obtains 
 \begin{equation}
  \tilde{m} = \tilde{E}x^{(1-\alpha)/\beta}[1 + \bar{u}_{4}x^{\theta_{4}/\beta} \pm \bar{u}_{1}x^{(\Delta -1)/\beta} \pm \bar{u}_{2}x^{(1-\alpha)/\beta} + \cdots],  \label{eq:3.56}
 \end{equation}
where the leading coefficients are
 \begin{eqnarray}
  \bar{u}_{1} & = & \tilde{u}_{1}-(1-\alpha)(E/V_{1}+u_{1})/\beta, \nonumber \\
  \bar{u}_{2} & = & \tilde{V}_{1}/\tilde{E}, \hspace{.2in} \bar{u}_{4} = \tilde{u}_{4} - (1-\alpha)u_{4}/\beta,  \label{eq:3.57}
 \end{eqnarray}
while the correction factor exponents have $d=3$ Ising values $\theta_{4}/\beta \simeq 1.64, \hspace{.2cm} (\Delta -1)/\beta \simeq 1.73\hspace{.2cm}$and$\hspace{.2cm}(1-\alpha)/\beta \simeq 2.73$.

\subsection{Binodals above the Endpoint Temperature}
\label{sec3.f}
Let us consider, first, the {\em spectator-phase binodal} ${\cal B}^{\alpha}$ above $T_{e}$ --- see Figs.~\ref{fig3}(c) and~\ref{fig6}(c) --- which is the simplest to analyze. Since $G^{\alpha}(g,t,h)$ is noncritical, the densities $m$ and $\tilde{m}$ are {\em noncritical} functions of $g$, $t$, and $h$ in the spectator-phase $\alpha$. At fixed $t > 0$, the phase boundary $g_{\sigma}(t;h)$ is also a nonsingular function of $h$ with $t$-dependent expansion coefficients, which are discussed explicitly below in Sec.~\ref{sec4}.E. Consequently, on the isotherms above $T_{e}$, the $\alpha$ binodal becomes noncritical. However, singularities of the binodal are to be expected as $T \rightarrow T_{e}+$.

In the {\em symmetric case}, by using the previous definitions (\ref{eq:3.34}) and (\ref{eq:3.35}) and the phase boundary $g_{\sigma}(t;h)$ given below in (\ref{eq:4.68}), we obtain
 \begin{equation}
  \tilde{m} = \tilde{m}_{e}^{\alpha} - \tilde{\chi}_{e}^{\alpha}(g_{\sigma,0}^{+}t + g_{\sigma,1}^{+}t^{2-\alpha}) - \tilde{\chi}_{e}^{\alpha}g_{\sigma,3}^{+}t^{-\gamma}x_{\alpha}^{2} + \cdots,  \label{eq:3.58}
 \end{equation}
where $x_{\alpha} = m/\chi_{e}^{\alpha}$, as for the symmetric noncritical endpoint binodals in (\ref{eq:3.36}), while the coefficients, $g_{\sigma,0}^{+}$, etc., are given below in (\ref{eq:4.69}). Note that the curvature of the binodal diverges like $t^{-\gamma}$ when $T \rightarrow T_{e}+$.

In the {\em nonsymmetric case}, using (\ref{eq:4.71}), we obtain
 \begin{equation}
  \tilde{m} = \tilde{m}_{e}^{\alpha} + \tilde{m}_{1}^{\alpha}x_{\alpha} + \tilde{m}_{2}^{\alpha}(t)x_{\alpha}^{2} + \cdots,   \label{eq:3.59}
 \end{equation}
where $\tilde{m}_{e}$ and $\tilde{m}_{1}^{\alpha}$ are given above in (\ref{eq:3.43}) and (\ref{eq:3.45}), respectively, while the coefficient of second order in $x_{\alpha}$ ($=m/\chi_{e}^{\alpha}$) is
 \begin{equation}
  \tilde{m}_{2}^{\alpha}(t) = 2(G_{4}^{\alpha} - L_{\rho}G_{5}^{\alpha})g_{\sigma,3}^{+}j_{1}^{2} t^{-\gamma} + \cdots,  \label{eq:3.60}
 \end{equation}
where $g_{\sigma,3}^{+}$ is given below in (\ref{eq:4.72}). Here we have neglected higher order corrections in $t$. Just as in the symmetric case, the curvature of the binodal diverges when $T \rightarrow T_{e}+$.

Consider next the {\em critical phase binodal} ${\cal B}^{\beta\gamma}$ above $T_{e}$: see Figs.~\ref{fig3}(c) and~\ref{fig6}(c). This may be determined using (\ref{eq:4.16}) below and its twin for $\tilde{m}$ with the aid of the spectator-phase boundary, $g_{\sigma}(t;\tilde{h})$, which is derived in Sec.~\ref{sec4}.E. For fixed $t>0$, the small $y$ expansion for the scaling function $W_{+}(y,y_{4},\cdots)$ yields only integer powers of $\tilde{h}$ in (\ref{eq:4.16}) and its twin so that the densities $m$ and $\tilde{m}$ are noncritical functions of $\tilde{h}$. Consequently, the critical phase binodal is again noncritical above $T_{e}$.

In the {\em symmetric case}, the densities $m$ and $\tilde{m}$ can be written in terms of $\tilde{h}$ by using (\ref{eq:4.16}) and its twin as
 \begin{eqnarray}
  m & = & l_{1}t^{-\gamma}\tilde{h} + \cdots,   \label{eq:3.61} \\
  \tilde{m} & = & \tilde{m}_{0}(t) + \tilde{l}_{2}t^{-\gamma-1}\tilde{h}^{2} + \cdots,  \label{eq:3.62}
 \end{eqnarray}
where $\tilde{m}_{0}(t)$ is a function of $t$ only while the coefficients are
 \begin{eqnarray}
  l_{1} & = & 2Q_{e}U^{2}W_{+2}^{0}|1-q_{1}(D_{2}/D_{1})|^{-\gamma},  \nonumber \\
  \tilde{l}_{2} & = & -\gamma q_{1}Q_{e}U^{2}W_{+2}^{0}|1-q_{1}(D_{1}/D_{2})|^{-\gamma-1}.  \label{eq:3.63}
 \end{eqnarray}
Notice that $\tilde{l}_{2}$ is negative in case {\bf A} while it is positive in case {\bf B}, as for $\tilde{A}$, the leading amplitude of the lambda line binodals: see the paragraph below (\ref{eq:3.9}). We may eliminate $\tilde{h}$ between (\ref{eq:3.61}) and (\ref{eq:3.62}) and write the binodal in terms of $x = m/l_{1}$, noticing $l_{1} > 0$, as
 \begin{equation}
  \tilde{m} = \tilde{m}_{0}(t) + \tilde{l}_{2}t^{\gamma -1}x^{2} + \cdots.  \label{eq:3.64}
 \end{equation}
Since $\gamma > 1$ in the $d < 4$ Ising universality classes, the coefficient of the quadratic term in $x$ vanishes as $T \rightarrow T_{e}+$. This result could be anticipated, since the critical endpoint binodals have the leading exponent $(1-\alpha)/\beta$ $(\simeq 2.73)$ in the symmetric case. Thus the curvature of ${\cal B}^{\beta\gamma}$ is singular but {\em non}divergent when $T \rightarrow T_{e}$.

In the general {\em nonsymmetric case}, the situation is more complicated. The densities can now be expressed as
 \begin{eqnarray}
  m & = & m_{0}(t) + l_{1}t^{-\gamma}\tilde{h} + l_{2}t^{-\gamma-1}\tilde{h}^{2} + \cdots,  \label{eq:3.65}  \\
  \tilde{m} & = & \tilde{m}_{0}(t) + \tilde{l}_{1}t^{1-\alpha}\tilde{h} + \tilde{l}_{2}t^{-\gamma-1}\tilde{h}^{2} + \cdots,  \label{eq:3.66}
 \end{eqnarray}
where the constant coefficients are presented below in (\ref{eq:4.74-1}). Note that the term linear in $\tilde{h}$ for $\tilde{m}$ has a leading $t$-dependent coefficient that vanishes when $T \rightarrow T_{e}+$. As before, the critical phase binodal can be written in terms of $x = \Delta m/l_{1}$ with $\Delta m \equiv m - m_{0}(t)$ as
 \begin{equation}
  \tilde{m} = \tilde{m}_{0}(t) + \tilde{l}_{1}t^{1+2\gamma -\alpha} x + \tilde{l}_{2}t^{\gamma -1} x^{2} + \cdots.   \label{eq:3.68}
 \end{equation}
Evidently, both the coefficients of $x$ and $x^{2}$ are singular but vanish when $T \rightarrow T_{e}+$ and $\gamma > 1$.

\subsection{Binodals below the Endpoint Temperature}
\label{sec3.g}
Below the endpoint temperature three phases, $\alpha$, $\beta$, and $\gamma$ may coexist on the triple line $\tau$. The binodals near a triple point then spring from the corners of a three-phase triangle. The corresponding phase diagrams in the density plane are shown in Figs.~\ref{fig3}(a) and~\ref{fig6}(a) for the two cases {\bf NA} and {\bf SB}, respectively. Thermodynamic stability then requires that these diagrams must satisfy Schreinemakers' rules: \cite{ref4,mef:yck,sch,jcw} details are given in Ref. 17.

The explicit forms of the {\em spectator-phase binodals}, ${\cal B}_{<}^{\alpha\pm}$, can be obtained without difficulty by using the phase boundary $g_{\sigma}(t,h)$ below $T_{e}$ as presented in (\ref{eq:4.76}) and (\ref{eq:4.79}) for the symmetric and nonsymmetric cases, respectively. In the {\em symmetric case}, the binodal may be expressed as
 \begin{equation}
  \tilde{m} = \tilde{m}_{e}^{\alpha} - \tilde{\chi}_{e}^{\alpha}g_{\sigma,0}^{-}t \mp \tilde{\chi}_{e}^{\alpha}g_{\sigma,2}^{-}|t|^{\beta}x_{\alpha} - \tilde{\chi}_{e}^{\alpha}g_{\sigma,3}^{-}|t|^{-\gamma}x_{\alpha}^{2} + \cdots,   \label{eq:3.69}
 \end{equation}
where $x_{\alpha} = m/\chi_{e}^{\alpha}$ and the upper (lower) sign corresponds to $m > 0$ ($<0$), while the coefficients, $g_{\sigma,0}^{-}$, etc., are given below in (\ref{eq:4.77}). Note that the slope vanishes as $T \rightarrow T_{e}-$ while the curvature diverges as $|t|^{-\gamma}$. In the {\em nonsymmetric case}, the binodal is given by
 \begin{equation}
  \tilde{m} = \tilde{m}_{e}^{\alpha} + \tilde{m}_{1}^{\alpha}(t) x_{\alpha} + \tilde{m}_{2}^{\alpha}(t) x_{\alpha}^{2} + \cdots,   \label{eq:3.70}
 \end{equation}
where the coefficients are
 \begin{eqnarray}
  \tilde{m}_{1}^{\alpha}(t) & = & \tilde{m}_{1}^{\alpha} \mp 2 g_{\sigma,2}^{-}j_{1}(-G_{4}^{\alpha} + L_{\rho}G_{5}^{\alpha})|t|^{\beta} + \cdots,   \nonumber \\
  \tilde{m}_{2}^{\alpha}(t) & = & -2 g_{\sigma,3}^{-}j_{1}^{2}(-G_{4}^{\alpha} + L_{\rho}G_{5}^{\alpha})|t|^{-\gamma} + \cdots,     \label{eq:3.71}
 \end{eqnarray}
while the $\mp$ signs again correspond to $m \gtrless 0$. The coefficients, $g_{\sigma,0}^{-}$, etc., are given below in (\ref{eq:4.80}). Notice that the linear terms do not vanish, but approach the same value when $T \rightarrow T_{e}-$.

The {\em critical phase binodals}, ${\cal B}_{<}^{\beta}$ and ${\cal B}_{<}^{\gamma}$, can be obtained, in principle, by using (\ref{eq:4.16}) and its twin with the aid of the phase boundary $g_{\sigma}(t,h)$ given below in (\ref{eq:4.76}) and (\ref{eq:4.79}). However, the analysis becomes more complicated, since these binodals are associated with the lambda line binodals near the vertices of the three-phase triangle: see Figs.~\ref{fig3}(a) and~\ref{fig6}(a). Hence, we do not present their explicit forms here. One can anticipate, however, that the binodals have linear slopes and quadratic terms which both vanish when $T \rightarrow T_{e}-$ in the $d < 4$ Ising universality classes. 
 
\section{Derivation of the Binodal Expressions}
\label{sec4}
In this section we sketch, for completeness, some of the details entering the derivation of the results for the binodals presented in Sec.~\ref{sec3} from the postulates of Sec.~\ref{sec2}. In addition, we give explicit expressions for the leading amplitudes entering the formulae of Sec.~\ref{sec3} in terms of the original parameters of the postulated free energies of Sec.~\ref{sec2}.

\subsection{Principles for Obtaining Isothermal Sections}
\label{sec4.a}
Our aim is to describe isothermal sections of the full $(g,t,h)$ phase space in terms of the density variables
 \begin{equation}
  \rho_{1} = -\partial_{h}G, \hspace{.2in} \rho_{2} = -\partial_{g}G \hspace{.2in} \mbox{with} \hspace{.2in} \partial_{h} \equiv \partial/\partial h, \hspace{.2in} \partial_{g} \equiv \partial/\partial g. \label{eq:4.1}
 \end{equation}
Accordingly, we treat $t$ as a fixed parameter and regard only $g$ and $h$ as varying. The basic nonlinear scaling fields $\tilde{t}$ and $\tilde{h}$ are then to be viewed as functions only of $g$ and $h$. Once the appropriate derivatives with respect to $g$ and $h$ have been performed, however, it is more convenient, in light of the scaling postulate (\ref{eq:2.13}), to employ the nonlinear scaling fields $\tilde{t}$ and $\tilde{h}$ as the primary field variables. Note, in particular, that both the $\lambda$ line and the triple line, $\tau$, lie in the plane $\tilde{h} = 0$. Beyond that, the $\lambda$-line or $\rho$-surface binodals also correspond to $\tilde{h} = 0$ while the spectator-phase and $\sigma$ binodals are of interest only for small $\tilde{h}$. Consequently we express $g$ and $h$ in terms of $\tilde{t}$ and $\tilde{h}$ via the noncritical expansions
 \begin{eqnarray}
  g & = & g_{\lambda}(t) + e_{1}\tilde{t} + e_{2}\tilde{h} + e_{3}\tilde{t}^{\,2} + e_{4}\tilde{t}\tilde{h} + e_{5}\tilde{h}^{2} + \cdots,  \label{eq:4.2}  \\
  h & = & h_{\lambda}(t) + f_{1}\tilde{t} + f_{2}\tilde{h} + f_{3}\tilde{t}^{\,2} + f_{4}\tilde{t}\tilde{h} + f_{5}\tilde{h}^{2} + \cdots,  \label{eq:4.3}
 \end{eqnarray}
where the $\lambda$-line values, $g_{\lambda}$ and $h_{\lambda}$, were introduced in (\ref{eq:2.8}) and seen to be noncritical functions. Likewise, all the coefficients $e_{j}(t)$ and $f_{j}(t)$ are noncritical with, in the symmetric case,
 \begin{eqnarray}
  \mbox{{\bf S}}: \hspace{.3in} &  & e_{2} = e_{4} = f_{1} = f_{3} = f_{5} = 0,  \nonumber \\
                  &  & e_{1} = q_{1}^{-1} + O(t),\, f_{2} = 1,\, e_{3}= -\frac{q_{2}}{q_{1}^{3}},\, e_{5} = -\frac{q_{6}}{q_{1}},\, f_{4} = -\frac{r_{1}}{q_{1}}.  \label{eq:4.4}
 \end{eqnarray}
More generally, with $\Lambda_{0} \equiv q_{1} - r_{0}q_{0}\, (\neq 0)$, we have
 \begin{equation}
  \mbox{{\bf N}}: \hspace{.3in} e_{1},\, e_{2},\, f_{1},\, f_{2} = (1,\, -q_{0},\, -r_{0},\, q_{1})/\Lambda_{0} + O(t),  \label{eq:4.5}
 \end{equation}
while $e_{3},\cdots,f_{5}$ are also readily found in terms of the $q_{j}$ and $r_{j}$.

Any noncritical property ${\P}(g,t,h)$ with expansion (\ref{eq:2.2}) can then be rewritten as
 \begin{equation}
  {\P}(g,t,h) = {\P}_{\lambda}(t) + \mbox{\boldmath $\dot{\hspace{-1.0mm}{\P}}$}_{1}(t)\tilde{t} + \mbox{\boldmath $\dot{\hspace{-1.0mm}{\P}}$}_{2}(t)\tilde{h} + \mbox{\boldmath $\dot{\hspace{-1.0mm}{\P}}$}_{3}(t)\tilde{t}^{\,2} + \cdots,  \label{eq:4.6}
 \end{equation}
where the value on the $\lambda$ line is given by
 \begin{eqnarray}
  &  &\hspace{.3in} {\P}_{\lambda}(t) = {\P}_{e} + {\P}_{\lambda 1}t + {\P}_{\lambda 2}t^{2} + \cdots,  \label{eq:4.7} \\
  {\P}_{\lambda 1} & = & {\P}_{1}\Lambda_{g} + {\P}_{2} + {\P}_{3}\Lambda_{h}, \hspace{.3in} {\P}_{\lambda 2} = {\P}_{1}\Lambda_{g2} + \cdots + {\P}_{9}\Lambda_{h}^{2},\, \cdots,  \label{eq:4.8}
 \end{eqnarray}
where $\Lambda_{g}$, $\Lambda_{h}$, $\Lambda_{g2}$, etc.$\hspace{.1cm}$are defined in (\ref{eq:2.8}) and (\ref{eq:2.9}), while the remaining noncritical coefficients take the form
 \begin{eqnarray}
  \mbox{\boldmath $\dot{\hspace{-1.0mm}{\P}}$}_{j}(t) & = & \mbox{\boldmath $\dot{\hspace{-1.0mm}{\P}}$}_{je} + \mbox{\boldmath $\dot{\hspace{-1.0mm}{\P}}$}_{j1}t + \mbox{\boldmath $\dot{\hspace{-1.0mm}{\P}}$}_{j2}t^{2} + \cdots,\hspace{.2in} \mbox{\boldmath $\dot{\hspace{-1.0mm}{\P}}$}_{je} = {\P}_1 e_{j} + {\P}_{3}f_{j},  \label{eq:4.9}  \\
  \mbox{\boldmath $\dot{\hspace{-1.0mm}{\P}}$}_{j1} & = & 2 ({\P} _{4}\Lambda_{g}e_{j} + {\P}_{5}\Lambda_{g}f_{j} + {\P}_{5}\Lambda_{h}e_{j} + {\P}_{6}e_{j} + {\P}_{7}f_{j} + {\P}_{9}\Lambda_{h}f_{j}),  \label{eq:4.10}
 \end{eqnarray}
for $j=1$ or $2$, and
 \begin{equation}
  \mbox{\boldmath $\dot{\hspace{-1.0mm}{\P}}$}_{3e} = {\P}_{1}e_{3} + {\P}_{3}f_{3} + {\P}_{4}e_{1}^{2} +2{\P}_{5}e_{1}f_{1} + {\P}_{9}f_{1}^{2},  \label{eq:4.11}
 \end{equation}
and so on.

Of course, we eventually wish to eliminate $\tilde{t}$ and $\tilde{h}$ in favor of $\rho_{1}$ and $\rho_{2}$ or, in view of the discussion of Sec.~\ref{sec3}.A in terms of 
 \begin{equation}
  m = -\partial G + (\partial G)_{e} \hspace{.2in} \mbox{and} \hspace{.2in}\tilde{m} = -\tilde{\partial}G + (\tilde{\partial}G)_{e},  \label{eq:4.12}
 \end{equation}
where the compound differential operators are 
 \begin{equation}
  \partial = \partial_{h} - L_{\sigma}\partial_{g} \hspace{.2in} \mbox{and} \hspace{.2in} \tilde{\partial} = \partial_{g} - L_{\rho}\partial_{h}.   \label{eq:4.13}
 \end{equation}
However, once we have expressions for $m$ and $\tilde{m}$ in terms of $\tilde{t}$ and $\tilde{h}$ we can regard these fields merely as auxiliary {\em parameters} relating $m$ and $\tilde{m}$. Note in particular, that coexisting phases must have the same values of $\tilde{t}$ and $\tilde{h}$. Thus for the $\rho$ binodals we can put $s = (-\tilde{t})^{\beta}$, for $\tilde{t} < 0$, and set $\tilde{h} = 0$. This indicates the origin of the parametric descriptions of the binodals presented in Sec.~\ref{sec3}.B. Similarly, for the binodals associated with the $\sigma$ phase boundary, equating the free energies $G^{\beta\gamma}$ and $G^{\alpha}$ gives a relation for $\tilde{t}$ in terms of $\tilde{h}$ (and $t$); then $\tilde{h}$ is an appropriate parameter.

The axis slopes $L_{\sigma}$ and $L_{\rho}$ in (\ref{eq:4.13}) were explained in Sec.~\ref{sec3}.A and the slope $L_{\sigma}$ was given in (\ref{eq:3.42}). Below we will establish the $t$-dependent result
 \begin{equation}
  L_{\rho}(t) = r_{0} + [ 2r_{4}\Lambda_{g} + r_{5} + r_{1}\Lambda_{h} - r_{0}(r_{1}\Lambda_{g} + r_{2} + 2r_{3}\Lambda_{h})]t + O(t^{2}),  \label{eq:4.15}
 \end{equation}
where $L_{\rho}(t)$ was introduced just before (\ref{eq:3.13}) and $L_{\rho} \equiv L_{\rho}(0) = r_{0}$.

Now using (\ref{eq:4.12}) and (\ref{eq:2.13}) we obtain the primary density in the form
 \begin{eqnarray}
  m & = & (\partial G^{0})_{e} - \partial G^{0} + |\tilde{t}|^{2-\alpha}\left[ (\partial Q){\mathit{W}}_{\pm} + \sum_{k \geq 4} (\partial U_{k})Q{\mathit{W}}_{\pm}^{\prime (k)}|\tilde{t}|^{\theta_{k}}\right]  \nonumber \\
    &   & \pm (\partial\tilde{t})|\tilde{t}|^{1-\alpha}Q\mbox{\boldmath $\dot{\hspace{-1.0mm}{\mathit{W}}}$}_{\pm} \mp (\partial\tilde{t})\tilde{h}|\tilde{t}|^{\beta-1}\Delta QU{\mathit{W}}_{\pm}^{\prime} + (\partial\tilde{h})|\tilde{t}|^{\beta}QU{\mathit{W}}_{\pm}^{\prime},  \label{eq:4.16}
 \end{eqnarray}
for $\tilde{h} \rightarrow 0$, with a precisely similar expression for $\tilde{m}$ with $\tilde{\partial}$ replacing $\partial$, while
 \begin{eqnarray}
  \mbox{\boldmath $\dot{\hspace{-1.0mm}{\mathit{W}}}$}_{\pm}(y, y_{4}, \cdots) & = & (2-\alpha){\mathit{W}}_{\pm} + \sum_{k \geq 4} \theta_{k}U_{k}{\mathit{W}}_{\pm}^{\prime (k)}|\tilde{t}|^{\theta_{k}},  \label{eq:4.17}  \\
  {\mathit{W}}_{\pm}^{\prime} (y, y_{4}, \cdots) & = & (\partial {\mathit{W}}_{\pm}/\partial y), \hspace{.2in} {\mathit{W}}_{\pm}^{\prime (k)}(y, \cdots) = (\partial {\mathit{W}}_{\pm}/\partial y_{k}).  \label{eq:4.18}
 \end{eqnarray}
Note that $\partial G^{0}$ and the coefficients $\partial Q$, $\partial U_{k}$, $\partial\tilde{t}$, and $\partial\tilde{h}$ are all noncritical and so can be written as in (\ref{eq:4.6}). This form thus enables one to identify all the singular terms appearing in $m$ and $\tilde{m}$.

Now on the $\lambda$ line we have $\tilde{t} = \tilde{h} = 0$. Thus (\ref{eq:4.16}) and its twin for $\tilde{m}$ yield the expansions (\ref{eq:3.4}) and (\ref{eq:3.5}) for $m_{\lambda}$ and $\tilde{m}_{\lambda}$ with
 \begin{eqnarray}
  M_{1} & = & 2\left[ L_{\sigma}(\Lambda_{g}G^{0}_{4} + \Lambda_{h}G^{0}_{5} + G^{0}_{6}) - \Lambda_{g}G^{0}_{5} - G^{0}_{7} - \Lambda_{h}G^{0}_{9}\right],  \label{eq:4.19} \\
  \tilde{M}_{1} & = & 2\left[ L_{\rho}(\Lambda_{g}G^{0}_{5} + G^{0}_{7} + \Lambda_{h}G^{0}_{9}) - \Lambda_{g}G^{0}_{4} - \Lambda_{h}G^{0}_{5} - G^{0}_{6}\right],  \label{eq:4.20}
 \end{eqnarray}
so that $M_{1} = 0$ and $\tilde{M}_{1} = 2(G^{0}_{4}/q_{1} - G^{0}_{6})$ in the symmetric case. Defining $R(g,t,h) = (\partial G^{0})_{e} - \partial G^{0}$ and $\tilde{R}$ likewise, and expanding as in (\ref{eq:4.6}) yields, for $j=1, 2$,
 \begin{eqnarray}
  \mbox{\boldmath $\dot{\hspace{-1.0mm}\mathit{R}}$}_{j} & = & 2\left[ L_{\sigma}(G^{0}_{4}e_{j} + G^{0}_{5}f_{j}) - G^{0}_{5}e_{j} - G^{0}_{9}f_{j}\right] + O(t),  \label{eq:4.21}  \\
  \mbox{\boldmath $\dot{\hspace{-1.0mm}\mathit{\tilde{R}}}$}_{j} & = & 2\left[ r_{0}(G^{0}_{5}e_{j} + G^{0}_{9}f_{j}) - G^{0}_{4}e_{j} - G^{0}_{5}f_{j}\right] + O(t),  \label{eq:4.22}
 \end{eqnarray}
where (\ref{eq:4.15}) was used for $L_{\rho}$. For reference below we also record
 \begin{eqnarray}
  (\partial\tilde{t})_{\lambda} & = & q_{0} - L_{\sigma}q_{1} + \left[q_{5}\Lambda_{g} + 2q_{6}\Lambda_{h} + q_{7} - L_{\sigma}(2q_{2}\Lambda_{g} + q_{3} + q_{5}\Lambda_{h})\right]t + \cdots,  \label{eq:4.23} \\
  (\partial\tilde{h})_{\lambda} & = & 1 - L_{\sigma}r_{0} + \left[ r_{1}\Lambda_{g} + r_{2} +2r_{3}\Lambda_{h} - L_{\sigma}(r_{1}\Lambda_{h} + 2r_{4}\Lambda_{g} +r_{5})\right]t + \cdots,  \label{eq:4.24} \\
  (\tilde{\partial}\tilde{t})_{\lambda} & = & q_{1} - r_{0}q_{0} + \left[ 2q_{2}\Lambda_{g} + q_{3} + q_{5}\Lambda_{h} - L_{\rho}(q_{5}\Lambda_{g} + 2q_{6}\Lambda_{7} + q_{7})\right]t + \cdots,  \label{eq:4.25}  \\
  (\tilde{\partial}\tilde{h})_{\lambda} & = & r_{0} - L_{\rho} + \left[ r_{1}\Lambda_{h} + 2r_{4}\Lambda_{g} + r_{5} - L_{\rho}(r_{1}\Lambda_{g} + r_{2} + 2r_{3}\Lambda_{h}) \right]t + \cdots.  \label{eq:4.26}
 \end{eqnarray}
Clearly, any desired higher order terms in the $\tilde{t}$, $\tilde{h}$ expansions can be obtained straightforwardly. Finally, we remark that we will shortly see that the condition determining $L_{\rho}(t)$ is that $(\tilde{\partial}\tilde{h})_{\lambda}$ vanishes term by term; substitution of (\ref{eq:4.15}) in (\ref{eq:4.26}) checks this.

\subsection{Derivation of the $\lambda$-Line Binodals}
\label{sec4.b}
The binodals associated with the $\lambda$ line may, essentially, be obtained directly from (\ref{eq:4.16}) and its twin, by letting $\tilde{h} \rightarrow 0\pm$ with $\tilde{t} < 0$. In doing this the small $y$ expansions (\ref{eq:2.21}) must be used with attention to the $\sigma_{\boldkappa}(y)$ factors defined in (\ref{eq:2.22}). When this is done the $G^{0}$ terms in (\ref{eq:4.16}) generate only integral powers of $|\tilde{t}|$; the terms in $|\tilde{t}|^{2-\alpha}$ act merely to modify the correction factor of the $|\tilde{t}|^{1-\alpha}$ term. Note that the $\tilde{t}$ and $\tilde{h}$ expansions of $Q$ and of the $U_{k}$ yield correction terms varying as $|\tilde{t}|^{n+\theta[\boldkappa]}$ for all integers $n \geq 0$ and all $\boldkappa > 0$. The term in $\tilde{h}|\tilde{t}|^{\beta -1}$, which diverges as $|\tilde{t}| \rightarrow 0$, vanishes identically. Lastly, the term in $|\tilde{t}|^{\beta}$ contributes both to $m$ and $\tilde{m}$.

Introducing the parameter $s = |\tilde{t}|^{\beta}$ then yields the previously quoted expansions (\ref{eq:3.6}) and (\ref{eq:3.7}) for $m$ and $\tilde{m}$ in the symmetric case. The linear term in $s$ is absent in this $\tilde{m}$ expansion because the coefficient $(\partial_{g}\tilde{h})$ vanishes identically by symmetry when $h \equiv \tilde{h} \rightarrow 0$ and $L_{\rho} = 0$ is dictated. Similarly, terms varying as $s^{1/\beta}$ and $s^{(1-\alpha)/\beta}$ are absent in the expression for $m$ since $L_{\sigma} = 0$ and thence $\partial G^{0}$, $\partial Q$ and $\partial\tilde{t}$ all vanish. For {\em even} $k$ the derivatives $\partial_{h}U_{k} = \partial U_{k}$ (for $L_{\sigma} = h = 0$) also vanish by symmetry. However, in the fully symmetric situation each odd scaling field, $U_{2j+1}(g,t,h)$, must itself be odd in $h$: see (\ref{eq:2.12}). Hence after operating with $\partial_{h}$, contributions with {\em odd} $k$ in the terms proportional to $|\tilde{t}|^{2-\alpha+\theta_{k}}(\partial U_{k})$ in (\ref{eq:4.16}) appear in the expansion for $m$ in the symmetric case. Since $2 - \alpha = \beta + \Delta$ these terms are responsible for the appearance of the correction factors $|\tilde{t}|^{\Delta} = s^{\delta}$ in (\ref{eq:3.6}); see also (\ref{eq:3.9}). For completeness we record the leading amplitude values
 \begin{eqnarray}
  \tilde{A}_{e} & = & -(2-\alpha)q_{1}Q_{e}W_{-0}^{0}, \hspace{.1in} B_{e} = UQ_{e}W_{-1}^{0}, \hspace{.1in} \tilde{K}_{e} = 2G^{0}_{4}/q_{1},  \label{eq:4.27}  \\
  \tilde{a}_{4e} & = & \left( 1 + \frac{\theta_{4}}{2-\alpha}\right)\frac{W_{-0}^{(4)}}{W_{-0}^{0}}U_{4e}, \hspace{.2in} b_{4e} = \frac{W_{-1}^{(4)}}{W_{-1}^{0}}U_{4e}.  \label{eq:4.28}
 \end{eqnarray}
Clearly all other amplitudes are readily generated although their complexity increases rapidly with order.

In the general {\em nonsymmetric case} the $U_{k}$ for odd $k$ need not vanish when $\tilde{h} \rightarrow 0$ but the scaling function, $W_{-}(y, y_{4}, y_{5}, \cdots)$, still has special behavior for small $y_{k}$ when $k$ is odd: see (\ref{eq:2.18}). This is the reason why the $\pm$ signs (corresponding to $\tilde{h} \rightarrow 0\pm$) appear in the expansion (\ref{eq:3.11}) for $m$. The expansion for the secondary density $\tilde{m}$, when initially generated, has a similar structure. In particular, the leading term is proportional to $|\tilde{t}|^{\beta} \equiv s$. However, at this point we should, as explained in Sec.~\ref{sec3}.A, complete the specification of the density $\tilde{m}$ by appropriately choosing $L_{\rho}(t)$. This should be done by examining the common tangent to the critical binodals, namely ${\cal B}_{e}^{\beta}$ and ${\cal B}_{e}^{\gamma}$, at the endpoint: see Figs.~\ref{fig3}(b) and~\ref{fig6}(b). But these binodals involve the $\sigma$ phase boundary which we have not yet studied. Instead, we will select $L_{\rho}$ so that the common tangent of the $\lambda$-line binodals ${\cal B}_{e}^{\lambda +}$ and ${\cal B}_{e}^{\lambda -}$ or ${\cal B}_{>}^{\lambda +}$ and ${\cal B}_{>}^{\lambda -}$ coincides with the $\tilde{m} = 0$ axis when extrapolated to the endpoint. It will be confirmed below that this criterion gives the same value for $L_{\rho}$. The coefficient of the offending $|\tilde{t}|^{\beta}$ term is $(\tilde{\partial}\tilde{h})_{e}$: see (\ref{eq:4.26}). This vanishes when $L_{\rho} = r_{0}$ so confirming (\ref{eq:4.15}) for $t = 0$.

The residual $t$ and $|\tilde{t}|$ dependence of $(\tilde{\partial}\tilde{h})$ then yield the $B^{\prime} ts$ and $\tilde{B}s^{(1+\beta)/\beta}$ terms in the expansion (\ref{eq:3.12}) for $\tilde{m}$. The latter term is unavoidable in general and further complicates the singular corrections to the $\rho$ binodals in the nonsymmetric case. Nevertheless, as explained in Sec.~\ref{sec3}.B, the former term, linear in $s$, can be eliminated by adopting a temperature-dependent definition for $\tilde{m}$ by allowing $L_{\rho}$ to vary noncritically with $T$. The criterion now is to make $(\tilde{\partial}\tilde{h})_{\lambda}(t)$ vanish. Reference to (\ref{eq:4.26}) then confirms the leading term in $L_{\rho}(t)$ presented in (\ref{eq:4.15}).

The leading amplitudes in (\ref{eq:3.11}) and (\ref{eq:3.12}) for the nonsymmetric case are, recalling (\ref{eq:4.21})-(\ref{eq:4.26}) and (\ref{eq:4.28}),
 \begin{eqnarray}
  A(t), && \hspace{-.2in}\tilde{A}(t) = -(2-\alpha)Q_{\lambda}W_{-0}^{0}[ (\partial\tilde{t})_{\lambda}, (\tilde{\partial}\tilde{t})_{\lambda}],  \label{eq:4.29} \\
  B(t) & = & Q_{\lambda}UW_{-1}^{0}(\partial\tilde{h})_{\lambda},  \label{eq:4.30} \\
  \tilde{B}_{e} & = & -Q_{e}UW_{-1}^{0} [(2r_{4} - r_{0}r_{1})e_{1} + (r_{1} - 2r_{0}r_{3})f_{1}], \label{eq:4.31} \\
  K_{e} & = & -\mbox{\boldmath $\dot{\hspace{-1.0mm}\mathit{R}}$}_{1e},\hspace{.2in} \tilde{K}_{e} = -\mbox{\boldmath $\dot{\hspace{-1.0mm}\mathit{\tilde{R}}}$}_{1e},\hspace{.2in} \tilde{a}_{4} = a_{4},\hspace{.2in} \tilde{b}_{4} = b_{4}.  \label{eq:4.32}
 \end{eqnarray}
One further has $\tilde{a}_{5} = a_{5}$, $\tilde{b}_{5} = b_{5}$, etc., although correction terms carrying `noncritical factors' $s^{1/\beta} \equiv |\tilde{t}|$ do {\em not} in general satisfy corresponding equalities.

\subsection{Spectator Phase Boundary: Endpoint Isotherm}
\label{sec4.c}
As indicated in Sec.~\ref{sec3}.C, the first step in studying the binodals not associated with the $\lambda$ line is to obtain the phase boundary $\sigma$ as specified by $g_{\sigma}(t,h)$. On recalling (\ref{eq:2.26}) and (\ref{eq:2.13}), one sees this is to be found by solving
 \begin{equation}
  D(g,t,h) = -Q|\tilde{t}|^{2-\alpha} W_{\pm}(y, y_{4}, \cdots),  \label{eq:4.33}
 \end{equation}
where $D(g,t,h)$ is noncritical with $D_{e} = 0$ and $D_{1}> 0$. Here we focus only on the endpoint isotherm, $T = T_{e}$ or $t = 0$. Now consider the argument $y$ in leading order, using (\ref{eq:2.5}) and (\ref{eq:2.6}):
 \begin{equation}
  y = U\tilde{h}/|\tilde{t}|^{\Delta} \approx U(h + r_{0}g)/|q_{0}h + q_{1}g|^{\Delta}.  \label{eq:4.34}
 \end{equation}
If $r_{0}$, $q_{0}$ and $q_{1}$ do not vanish (as in the generic {\em non}symmetric case) it is evident that when $g$, $h \rightarrow 0$ on $\sigma$ one in general has $y \sim [\max(|g|, |h|)]^{1-\Delta}$ which diverges to $\infty$ since $\Delta > 1$. Thus to study (\ref{eq:4.33}) on the endpoint isotherm we must utilize the large $y$ expansions (\ref{eq:2.23}) for the scaling functions entering (\ref{eq:2.15}). In the symmetric case one actually has $r_{0} = q_{0} = 0$; but it then transpires, as shown below, that $g_{\sigma} \sim |h|^{(2-\alpha)/\Delta}$ so that $y \sim |h|^{\alpha -1}$. Since $\alpha < 1$ we see that $y$ again diverges. Thus in (\ref{eq:4.33}) we must always use the expansion
 \begin{eqnarray}
  W_{\pm} & = & W_{\infty}^{0}|y|^{(2-\alpha)/\Delta}(1 \pm w_{1}^{0}|y|^{-1/\Delta} + w_{2}^{0}|y|^{-2/\Delta} \pm \cdots)  \nonumber \\
          &   & + W_{\infty}^{(4)}y_{4}|y|^{(2-\alpha+\theta_{4})/\Delta}(1 \pm w_{1}^{(4)}|y|^{-1/\Delta} + \cdots)  \nonumber \\
          &   & + W_{\infty}^{(5)}y_{5}\mbox{sgn}(y)|y|^{(2-\alpha+\theta_{5})/\Delta}(1 \pm w_{1}^{(5)}|y|^{-1/\Delta} + \cdots)  \nonumber \\
          &   & + \cdots,              \label{eq:4.35}
 \end{eqnarray}
where the $\pm$ signs correspond to $\tilde{t} \gtrless 0$.

The analysis is considerably simpler if one uses $\tilde{h}$ as a variable in place of $h$. To this end we rearrange (\ref{eq:2.5}) and (\ref{eq:2.6}) with $t = 0$ to obtain
 \begin{equation}
  h = \tilde{h} -r_{0}g - (r_{1}-2r_{0}r_{3})g\tilde{h} - r_{3}\tilde{h}^{2} - \bar{r}_{4}g^{2} + \cdots,  \label{eq:4.36}
 \end{equation}
where $\bar{r}_{4} = r_{4} - r_{0}r_{1} + r_{3}r_{0}^{2}$, and 
 \begin{equation}
  \tilde{t} = q_{0}\tilde{h} + p_{1}g + p_{2}\tilde{h}^{2} + p_{3}g\tilde{h} + p_{4}g^{2} + \cdots,  \label{eq:4.37}
 \end{equation}
where the leading coefficients are
 \begin{eqnarray}
  p_{1} & = & q_{1} - q_{0}r_{0}, \hspace{1.in}  p_{2} = q_{6} - q_{0}r_{3},  \nonumber \\
  p_{3} & = & q_{5} - q_{0}r_{1} + 2q_{0}r_{0}r_{3} - 2q_{6}r_{0}, \hspace{.2in} p_{4} = q_{2} - q_{5}r_{0} + q_{6}r_{0}^{2} - q_{0}\bar{r}_{4}.  \label{eq:4.38}
 \end{eqnarray}
Note that in the symmetric case one has $q_{0} = q_{5} = q_{7} = 0$, $r_{0} = r_{3} = r_{4} = 0$ and so $p_{3} = 0$; we may suppose $p_{1} \neq 0$.

Now, combining these results for the {\em symmetric case} yields the asymptotic equation
 \begin{equation}
  D_{1}g = -D_{4}g^{2} - D_{9}\tilde{h}^{2} - (Q_{e} + Q_{1}g + \cdots)|U\tilde{h}|^{(\delta + 1)/\delta}Z - \cdots   \label{eq:4.39}
 \end{equation}
with scaling factor, from (\ref{eq:4.35}),
 \begin{eqnarray}
  Z & = & W_{\infty}^{0}[ 1 \pm w_{1}^{0}|y|^{-1/\Delta} + w_{2}^{0}|y|^{-2/\Delta} \pm \cdots]  \nonumber \\
    &   & + W_{\infty}^{(4)}U_{4}(g,0,h)|U\tilde{h}|^{\theta_{4}/\Delta}[ 1 \pm w_{1}^{(4)}|y|^{-1/\Delta} + \cdots]  \nonumber \\
    &   & + \mbox{sgn}(y)W_{\infty}^{(5)}U_{5}(g,0,h)|U\tilde{h}|^{\theta_{5}/\Delta}[ 1 \pm \cdots] + \cdots.  \label{eq:4.40}
 \end{eqnarray}
These equations are to be solved together with
 \begin{equation}
  |y|^{-1/\Delta} = \frac{|\tilde{t}|}{|U\tilde{h}|^{1/\Delta}} = \frac{|q_{1}g|}{|U\tilde{h}|^{1/\Delta}}\left[ 1 + \frac{q_{2}}{q_{1}}g + \frac{q_{6}}{q_{1}}\frac{\tilde{h}^{2}}{g} + \cdots\right], \label{eq:4.41}
 \end{equation}
to yield $g = g_{\sigma}(\tilde{h})$. This can be accomplished iteratively by noting that in leading order $g_{\sigma} \approx -J|\tilde{h}|^{(\delta +1)/\delta}$, where $J$ was defined in (\ref{eq:3.24}); however, care is called for!

One obtains the result quoted in (\ref{eq:3.22})-(\ref{eq:3.25}) which may be supplemented by
 \begin{eqnarray}
  J_{2} & = & D_{9}/D_{1}, \hspace{.3in} J_{4} = [(D_{4}/D_{1}) - (Q_{1}/Q_{e})]J^{2},  \label{eq:4.42} \\
  c_{3} & = & w_{1}^{0}(q_{6} - q_{1}J_{2})/U^{1/\Delta}, \hspace{.2in} c_{4} = W_{\infty}^{(4)}U_{4e}U^{\theta_{4}/\Delta}/W_{\infty}^{0},  \label{eq:4.43} \\
  c_{4}^{\prime} & = & c_{4}w_{1}^{(4)}|q_{1}|J/U^{1/\Delta}, \hspace{.2in} c_{5} = W_{\infty}^{(5)}U_{5,3}U^{\theta_{5}/\Delta}/W_{\infty}^{0},  \label{eq:4.44}
 \end{eqnarray}
where $U_{5,3}$ is the first nonzero expansion coefficient of $U_{5}(g,0,h) \approx U_{5,3}h$ in the symmetric case. The expression (\ref{eq:3.26}) for $\tilde{h}(h)$ on $\sigma$ follows from (\ref{eq:2.6}) and (\ref{eq:3.22}) by reversion.

The phase boundary in the {\em nonsymmetric case} follows in the analogous way but greater care is needed because of the increased number of nonvanishing and competing terms. Thus on the right side of (\ref{eq:4.39}) the new terms $-D_{3}\tilde{h}$ and $-2\bar{D}_{5}g\tilde{h}$ appear, where $\bar{D}_{5} = D_{5} - \frac{1}{2}D_{3}(r_{1} - 2r_{0}r_{3}) - D_{9}r_{0}$. The former term dominates and so in leading order one now finds
 \begin{equation}
  g_{\sigma} \approx -J_{1}\tilde{h} - J|\tilde{h}|^{(\delta +1)/\delta},  \label{eq:4.45}
 \end{equation}
where $J_{1}$ was defined in (\ref{eq:3.27}). This in turn yields the new behavior
 \begin{equation}
  |y|^{-1/\Delta} = \frac{|\bar{q}|}{U^{1/\Delta}}|\tilde{h}|^{1-(1/\Delta)}\left[ 1 - \tilde{\sigma}_{h}\frac{p_{1}}{\bar{q}}J|\tilde{h}|^{1/\delta} \mp \tilde{\sigma}_{h}\frac{p_{1}d_{1}}{\bar{q}}J|\tilde{h}|^{(1-\alpha)/\Delta} + \cdots\right],  \label{eq:4.46}
 \end{equation}
where $\bar{q} = q_{0} - p_{1}J_{1} = \tilde{q}/j_{1}$ while $\tilde{q}$, $j_{1}$, and $\tilde{\sigma}_{h}$ were defined in (\ref{eq:3.30}) and (\ref{eq:3.31}).

In this way one obtains the result (\ref{eq:3.28})-(\ref{eq:3.32}) which must be supplemented by new expressions for $J_{2}$ and $J_{3}$ while
 \begin{equation}
  d_{2} = w_{2}^{0}\bar{q}^{\,2}/U^{2/\Delta},\hspace{.2in} d_{2}^{\prime} = 2w_{2}^{0}p_{1}\bar{q}J/U^{2/\Delta},\hspace{.2in} d_{2}^{\prime\prime} = w_{2}^{0}p_{1}^{2}J^{2}/U^{2/\Delta}.  \label{eq:4.47}
 \end{equation}
The expressions for $d_{3}^{\prime}$, $d_{3}^{\prime\prime}$, are long and uninformative but we quote
 \begin{eqnarray}
  d_{3} & = & w_{3}^{0}|\bar{q}|^{3}/U^{3/\Delta},\hspace{.3in} d_{4} = c_{4},\hspace{.3in} d_{4}^{\prime} = w_{1}^{(4)}d_{4}|\bar{q}|/U^{1/\Delta},  \nonumber \\
  d_{4}^{\prime\prime} & = & w_{1}^{(4)}d_{4}p_{1}J/U^{1/\Delta},\hspace{.3in} d_{5} = W_{\infty}^{(5)}U_{5e}U^{\theta_{5}/\Delta}/W_{\infty}^{0},  \nonumber \\
  d_{5}^{\prime} & = & w_{1}^{(5)}d_{5}|\bar{q}|/U^{1/\Delta},\hspace{.3in} d_{5}^{\prime\prime} = w_{1}^{(5)}d_{5}p_{1}J/U^{1/\Delta}.   \label{eq:4.48}
 \end{eqnarray}
Finally the remaining coefficients in (\ref{eq:3.33}) are
 \begin{eqnarray}
  j^{\prime} & = & r_{0}jJ|j_{1}|^{(\delta +1)/\delta},\hspace{.3in} j^{\prime\prime} = r_{0}j_{1}d_{1}J|j_{1}|^{(3-2\alpha-\beta)/\Delta},  \nonumber \\
  j_{2} & = & j_{1}^{3}[ r_{0}J_{2} + (r_{1} - 2r_{0}r_{3})J_{1} - r_{3} - \bar{r}_{4}J_{1}^{2}].  \label{eq:4.49}
 \end{eqnarray}

\subsection{Derivation of the Critical Endpoint Binodals}
\label{sec4.d}
The critical phase binodals at the endpoint may be obtained from (\ref{eq:4.16}) and its twin using the endpoint isotherm, $g_{\sigma}(\tilde{h})$, obtained in the previous section. In order to do so, it is more convenient to rewrite (\ref{eq:4.16}) as
 \begin{eqnarray}
  m & = & (\partial G^{0})_{e}-(\partial G^{0}) + |\tilde{t}|^{2-\alpha}\left[ (\partial Q)W_{\pm} + \sum_{k\geq 4} (\partial U_{k})Q W_{\pm}^{'(k)}|\tilde{t}|^{\theta_{k}}\right] \nonumber \\
    &   & \pm (\partial \tilde{t})Q|\tilde{t}|^{1-\alpha}\tilde{W}_{\pm} + (\partial \tilde{h})UQ |\tilde{t}|^{\beta} W_{\pm}^{'},  \label{eq:4.50}
 \end{eqnarray}
and similarly for $\tilde{m}$ with $\tilde{\partial}$ replacing $\partial$, while
 \begin{equation}
  \tilde{\mathit{W}}_{\pm} = \mbox{\boldmath $\dot{\hspace{-1.0mm}\mathit{W}}$}_{\pm} - \Delta y \mathit{W}_{\pm}^{'},  \label{eq:4.51}
 \end{equation}
where $\Delta = 2- \alpha - \beta$ has been used. At the critical endpoint, $t=0$, we use $\tilde{h}$ as an auxiliary parameter relating $m$ and $\tilde{m}$. Using (\ref{eq:4.36}) and (\ref{eq:4.37}), the noncritical functions, $(\partial G^{0})$, $(\partial Q)$, etc. can be expressed in terms of $\tilde{h}$. Recalling the general expansion (\ref{eq:2.2}) for a noncritical function $P(g,t,h)$, we find, for $t = 0$,
 \begin{equation}
  P(g,t=0,h) = P_{e} + P_{3}\tilde{h} + (P_{1}-r_{0}P_{3})g_{\sigma} + \cdots,  \label{eq:4.52}
 \end{equation}
and similarly for the derivatives
 \begin{eqnarray}
  \partial P & = & P_{3}-L_{\sigma}P_{1} + 2(P_{9}-L_{\sigma}P_{5})\tilde{h} + 2(P_{5}-L_{\sigma}P_{4}-r_{0}P_{9}+r_{0}L_{\sigma}P_{5})g_{\sigma} + \cdots,  \label{eq:4.53} \\
  \tilde{\partial} P & = & P_{1}-L_{\rho}P_{3} + 2(P_{5}-L_{\rho}P_{9})\tilde{h} + 2(P_{4}-L_{\rho}P_{5}-r_{0}P_{5}+r_{0}L_{\rho}P_{9})g_{\sigma} + \cdots.  \label{eq:4.54}
 \end{eqnarray}
Likewise, in terms of $g_{\sigma}(t=0,\tilde{h})$ we obtain
 \begin{eqnarray}
  \partial \tilde{t} & = & q_{0}-L_{\sigma}q_{1} + (2q_{6}-L_{\sigma}q_{5})\tilde{h} + (q_{5} - 2L_{\sigma}q_{2} - 2r_{0}q_{6} + L_{\sigma}r_{0}q_{5})g_{\sigma} + \cdots, \label{eq:4.55} \\
  \partial \tilde{h} & = & 1 - L_{\sigma}r_{0} + (2r_{3} - L_{\sigma}r_{1})\tilde{h} + (r_{1} - 2L_{\sigma}r_{4} - 2r_{0}r_{3} + L_{\sigma}r_{0}r_{1})g_{\sigma} + \cdots, \label{eq:4.56} \\
  \tilde{\partial} \tilde{t} & = & q_{1} - L_{\rho}q_{0} + (q_{5} - 2L_{\rho}q_{6})\tilde{h} + (2q_{2} - L_{\rho}q_{5} - r_{0}q_{5} + 2r_{0}L_{\rho}q_{6})g_{\sigma} + \cdots, \label{eq:4.57} \\
  \tilde{\partial} \tilde{h} & = & r_{0} - L_{\rho} + (r_{1} - 2L_{\rho}r_{3})\tilde{h} + (2r_{4} - L_{\rho}r_{1} - r_{0}r_{1} + 2r_{0}L_{\rho}r_{3})g_{\sigma} + \cdots. \label{eq:4.58}
 \end{eqnarray}

As discussed before, the argument $y$ of the scaling functions $W_{\pm}$ diverges to $\infty$ when the endpoint is approached on the $\sigma$ surface. Thus in (\ref{eq:4.50}) and its twin the large $y$ expansions (\ref{eq:2.23}) for the scaling functions must be used with attention to the $\sigma_{\boldkappa}(y)$ factors defined in (\ref{eq:2.22}) and the multiexponents $\theta[\boldkappa]$ in (\ref{eq:2.24}). When this is done, we finally obtain the critical endpoint binodals from (\ref{eq:4.50}).

Introducing the parameter $s=|\tilde{h}|^{\beta/\Delta}$ then yields the previously quoted expansions (\ref{eq:3.50}) and (\ref{eq:3.51}) for $m$ and $\tilde{m}$ in the {\em symmetric} case. The linear term in $s$ is absent in the expression for $\tilde{m}$ when we choose $L_{\rho}(0)=r_{0}$ which reinforces previous results. In the expression for $m$ the $G^{0}$ term in (\ref{eq:4.50}) provides a linear term in $\tilde{h}$ that yields the $s^{\Delta/\beta}$ term in (\ref{eq:3.50}); the terms in $|\tilde{t}|^{2-\alpha}$ provide the $s^{(2-\alpha+\Delta)/\beta}$ term and higher order corrections, since $(\partial Q)$ generates $\tilde{h}$ in leading order; the term in $|\tilde{t}|^{1-\alpha}$ provides the $s^{(1-\alpha+\Delta)/\beta}$ term for $(\partial \tilde{t})$ for the same reason; then, finally, the term in $|\tilde{t}|^{\beta}$ provides the leading $s$ behavior. In the expression for $\tilde{m}$ all the terms, except for one in $|\tilde{t}|^{1-\alpha}$, provide correction terms, $s^{(2-\alpha)/\beta}$, in (\ref{eq:3.51}); the leading behavior, $s^{(1-\alpha)/\beta}$, is generated by the term in $|\tilde{t}|^{1-\alpha}$.

The leading amplitudes are
 \begin{equation}
  E = \left[(2-\alpha)/\Delta\right] Q_{e} W_{\infty}^{0} U^{(2-\alpha)/\Delta}, \hspace{.3in} \tilde{E} = q_{1} Q_{e} W_{\infty}^{0} w_{1}^{0} U^{(1-\alpha)/\Delta}.  \label{eq:4.59}
 \end{equation}  
For the record, we also quote
 \begin{eqnarray}
  V_{1} & = & -2 G_{9}^{0}, \hspace{1.in} V_{2} = 2 q_{6} Q_{e} W_{\infty}^{0} w_{1}^{0} U^{(1-\alpha)/\Delta}, \nonumber \\
  V_{3} & = & 2 Q_{9} W_{\infty}^{0} U^{(2-\alpha)/\Delta}, \hspace{.3in} u_{4} = \frac{(2-\alpha+\theta_{4})}{(2-\alpha)}\frac{W_{\infty}^{(4)}}{W_{\infty}^{0}} U_{4e} U^{\theta_{4}/\Delta}, \nonumber \\
  u_{1} & = & -\frac{(1-\alpha)}{(2-\alpha)}w_{1}^{0}q_{1}J/U^{1/\Delta}, \hspace{.3in} \tilde{u}_{4} = \frac{W_{\infty}^{(4)}w_{1}^{(4)}}{W_{\infty}^{0}w_{1}^{0}} U_{4e} U^{\theta_{4}/\Delta}, \label{eq:4.60} \\
  \tilde{u}_{1} & = & -2\frac{w_{2}^{0}q_{1}J}{w_{1}^{0}U^{1/\Delta}}, \hspace{.1in} \tilde{V} = 2 G_{4}^{0} J + [Q_{1} + (2-\alpha)r_{1}Q_{e}/\Delta]W_{\infty}^{0}U^{(2-\alpha)/\Delta}. \nonumber
 \end{eqnarray}

In the general {\em nonsymmetric} case, the linear term in $s$ is still absent in the expression for $\tilde{m}$: see (\ref{eq:3.55}). The expression for $m$ in terms of $s$ is given in (\ref{eq:3.54}); the $G^{0}$ terms in (\ref{eq:4.50}) yield the $s^{\Delta/\beta}$ term, as in the symmetric case, while the terms in $|\tilde{t}|^{2-\alpha}$ provide the $s^{(2-\alpha)/\beta}$ term and that in $|\tilde{t}|^{1-\alpha}$ gives $s^{(1-\alpha)/\beta}$; the leading term, $s$, is still provided by the term in $|\tilde{t}|^{\beta}$. In the expression for $\tilde{m}$ the leading behavior is $s^{(1-\alpha)/\beta}$, as in the symmetric case, again provided by the $|\tilde{t}|^{1-\alpha}$ term; the $G^{0}$ term yields corrections of leading order $s^{\Delta/\beta}$, while the terms in $|\tilde{t}|^{2-\alpha}$ and $|\tilde{t}|^{\beta}$ give the $s^{(2-\alpha)/\beta}$ term in (\ref{eq:3.54}). The required amplitudes are
 \begin{eqnarray}
  E & = & \left[(2-\alpha)/\Delta\right] (1-L_{\sigma}r_{0}) Q_{e} W_{\infty}^{0} U^{(2-\alpha)/\Delta}, \nonumber \\
 \tilde{E} & = & (q_{1}-r_{0}q_{0}) Q_{e} W_{\infty}^{0} w_{1}^{0} U^{(1-\alpha)/\Delta}.
 \label{eq:4.61} 
 \end{eqnarray}
For the record, we also quote the correction amplitudes
 \begin{eqnarray}
  V_{1} & = & (q_{0} - L_{\sigma}q_{1}) Q_{e} W_{\infty}^{0}w_{1}^{0}U^{(1-\alpha)/\Delta}, \nonumber \\
  V_{2} & = & -2(G_{9}^{0} - L_{\sigma}G_{5}^{0}) + 2J_{1}(G_{5}^{0} - L_{\sigma}G_{4}^{0} - r_{0}G_{9}^{0} + r_{0}L_{\sigma}G_{5}^{0}),  \nonumber \\
  V_{3} & = & (Q_{3} - L_{\sigma}Q_{1}) W_{\infty}^{0} U^{(2-\alpha)/\Delta},  \label{eq:4.62} \\
  \tilde{V}_{1} & = & -2(G_{5}^{0} - r_{0}G_{9}^{0}) + 2J_{1}(G_{4}^{0} - 2r_{0}G_{5}^{0} + r_{0}^{2}G_{9}^{0}), \nonumber \\
  \tilde{V}_{2} & = & [Q_{1} + (2-\alpha)(r_{1} - 2r_{0}r_{3})Q_{e}/\Delta] W_{\infty}^{0} U^{(2-\alpha)/\Delta}, \nonumber
 \end{eqnarray}
and the leading further coefficients
 \begin{equation}
  u_{1} = \frac{(1-\alpha)}{(2-\alpha)}\bar{q}w_{1}^{0}/U^{1/\Delta}, \hspace{.3in} 
  v_{1} = \tilde{u}_{1} = 2\frac{w_{2}^{0}\bar{q}}{w_{1}^{0}U^{1/\Delta}}.  \label{eq:4.63}
 \end{equation}

\subsection{Spectator Phase Boundary: Isotherms above $T_{e}$}
\label{sec4.e}
In Sec.~\ref{sec4}.C, we studied the endpoint isothermal phase boundary, $g_{\sigma}(h)$, in order to discuss the endpoint binodals. By the same token we study the phase boundary $g_{\sigma}(t,h)$ above $T_{e}$ as the first step in determining the supercritical binodals. This boundary is found by equating the free energies, $G^{\alpha}(g,t,h)$ and $G^{\beta\gamma}(g,t,h)$, of the spectator and critical phases, respectively, which yields (\ref{eq:4.33}) with $\tilde{t}>0$. The extended triple line $\tilde{\tau}$ --- see Figs.~\ref{fig1} and~\ref{fig4} --- is defined by $\tilde{h} = 0$ for $\tilde{t} > 0$, implying $y = 0$. Since we consider only the vicinity of the extended triple line $\tilde{\tau}$, we must utilize the small $y$ expansion (\ref{eq:2.19}) for the scaling function $W_{+}(y,y_{4},y_{5},\cdots)$ in (\ref{eq:4.33}). Using $\tilde{h}$ as the principle variable, which is advantageous in discussing the critical phase binodal, ${\cal B}^{\beta\gamma}$, the scaling function $W_{+}(y,y_{4},y_{5},\cdots)$ can be expanded in integral powers of $\tilde{h}$ with $t$-dependent coefficients. The noncritical function $D(g,t,h)$ can be expanded similarly. Then, solving (\ref{eq:4.33}) for $g_{\sigma}(t;\tilde{h})$ yields the desired nonsingular expansion. In this section, we consider only the leading $t$-dependent behavior of the resulting coefficients.

Accordingly, we rearrange (\ref{eq:2.5}) and (\ref{eq:2.6}) for $t > 0$ using just the linear terms to obtain
 \begin{eqnarray}
  h & = & \tilde{h} - r_{-1}t - r_{0}g + \cdots,    \label{eq:4.64} \\
  \tilde{t} & = & (1 - q_{0}r_{-1})t + q_{0}\tilde{h} + (q_{1} - q_{0}r_{0})g + \cdots.  \label{eq:4.65}
 \end{eqnarray}
The higher order terms in (\ref{eq:2.5}) and (\ref{eq:2.6}) enter only as correction terms in the $t$-dependent coefficients. The noncritical function $D(g,t,h)$ is then expanded, by recalling (\ref{eq:2.2}) and $D_{e}=0$, as
 \begin{equation}
  D(g,t,h) = (D_{1} - r_{0}D_{3})g + (D_{2} - r_{-1}D_{3})t + D_{3}\tilde{h} + \cdots.  \label{eq:4.66}
 \end{equation}
Now we are in a position to find the isothermal boundary $g_{\sigma}(t;\tilde{h})$ above $T_{e}$.

In the {\em symmetric case}, we obtain
 \begin{equation}
  D_{1}g + D_{2}t + \cdots = -QW_{+0}^{0}|\tilde{t}|^{2-\alpha} - QW_{+2}^{0}U^{2}|\tilde{t}|^{-\gamma}\tilde{h}^{2} + \mbox{O}(\tilde{h}^{4}).    \label{eq:4.67}
 \end{equation}
By symmetry only even powers of $\tilde{h}$ appear. Solving for $g$ with the aid of (\ref{eq:4.65}) then yields
 \begin{equation}
  g_{\sigma}(t;\tilde{h}) = -g_{\sigma,0}^{+}t - g_{\sigma,1}^{+}t^{2-\alpha} - g_{\sigma,3}^{+}t^{-\gamma}\tilde{h}^{2} + \cdots,    \label{eq:4.68}
 \end{equation}
where the coefficients are
 \begin{eqnarray}
  g_{\sigma,0}^{+} & = & D_{2}/D_{1}, \hspace{.2in} g_{\sigma,1}^{+} = QD_{1}^{-3+\alpha}W_{+0}^{0}|D_{1} - q_{1}D_{2}|^{2-\alpha}, \nonumber \\
  g_{\sigma,3}^{+} & = & QD_{1}^{\gamma-1}W_{+2}^{0}U^{2}|D_{1} - q_{1}D_{2}|^{-\gamma}.   \label{eq:4.69}
 \end{eqnarray}
Notice that the coefficient of the quadratic term in $\tilde{h}$ diverges as $T \rightarrow T_{e}+$. In terms of $h$, which is advantageous in deriving the spectator-phase binodal, ${\cal B}^{\alpha}$, we obtain the same leading $t$-dependent coefficients for $g_{\sigma}(t;h)$.

In the {\em nonsymmetric case}, terms linear in $\tilde{h}$ appear in the expansion of the scaling function $W_{+}(y,y_{4},y_{5},\cdots)$ owing to the odd $\boldkappa$ exponents in (\ref{eq:2.19}). However, these terms only provide correction terms to the leading $t$-dependent behavior. Combining all the previous results yields the equation
 \begin{eqnarray}
 (D_{1}-r_{0}D_{3})g & + & (D_{2}-r_{-1}D_{3})t + D_{3}\tilde{h} + \cdots  \nonumber \\
  & = & -QW_{+0}^{0}|\tilde{t}|^{2-\alpha} - QW_{+2}^{0}U^{2}|\tilde{t}|^{-\gamma}\tilde{h}^{2} + \cdots.   \label{eq:4.70}
 \end{eqnarray}
Solving for $g$ yields
 \begin{equation}
  g_{\sigma}(t;\tilde{h}) = - g_{\sigma,0}^{+}t - g_{\sigma,1}^{+}t^{2-\alpha} - J_{1}\tilde{h} - g_{\sigma,3}^{+}t^{-\gamma}\tilde{h}^{2} + \cdots,    \label{eq:4.71}
 \end{equation}
where $J_{1}$ is given above in (\ref{eq:3.27}) while the other coefficients are
 \begin{eqnarray}
  g_{\sigma,0}^{+} & = & (D_{2}-r_{-1}D_{3})/(D_{1}-r_{0}D_{3}),   \nonumber  \\
  g_{\sigma,1}^{+} & = & \frac{QW_{+0}^{0}}{(D_{1}-r_{0}D_{3})}|\tilde{t}_{\sigma}|^{2-\alpha}, \hspace{.2in} g_{\sigma,3}^{+} = \frac{QW_{+2}^{0}U^{2}}{(D_{1}-r_{0}D_{3})}|\tilde{t}_{\sigma}|^{-\gamma},  \label{eq:4.72}
 \end{eqnarray}
in which the numerical factor is
 \begin{equation}
  \tilde{t}_{\sigma} = (1-q_{0}r_{-1}) - (q_{1}-q_{0}r_{0})[(D_{2}-r_{-1}D_{3})/(D_{1}-r_{0}D_{3})].   \label{eq:4.73}
 \end{equation}
Notice, again, that the coefficient of the quadratic term in $\tilde{h}$ diverges when $T\rightarrow T_{e}+$. The result (\ref{eq:4.71}) can be expanded in terms of $h$ by making the substitution
 \begin{equation}
  \tilde{h} = j_{1}h - r_{-1}t + \cdots,    \label{eq:4.74}
 \end{equation}
where $j_{1}$ is given in (\ref{eq:3.31}). By utilizing (\ref{eq:4.71}), the coefficients $l_{1}$, $\cdots$, $\tilde{l}_{2}$ in (\ref{eq:3.65}) and (\ref{eq:3.66}) are found to be
 \begin{eqnarray}
  l_{1} & = & 2(1 - L_{\sigma}r_{0})Q_{e}U^{2}W_{+2}^{0}|\tilde{t}_{\sigma}|^{-\gamma},  \nonumber  \\
  l_{2} & = & -\gamma (q_{0} - L_{\sigma}q_{1})Q_{e}U^{2}W_{+2}^{0}|\tilde{t}_{\sigma}|^{-\gamma -1}, \nonumber \\
  \tilde{l}_{1} & = & (2 - \alpha)W_{+0}^{0}|\tilde{t}_{\sigma}|^{1-\alpha} [\{ (q_{5} - 2r_{0}q_{6}) - 2J_{1}(q_{2} - r_{0}q_{5} + r_{0}^{2}q_{6})\} Q_{e} \nonumber \\
  &  & \hspace{1.in} + (q_{1} - r_{0}q_{0})\{ Q_{3} - J_{1}(Q_{1} - r_{0}Q_{3})\} ],  \label{eq:4.74-1} \\
  \tilde{l}_{2} & = & -\gamma (q_{1} - r_{0}q_{0})Q_{e}U^{2}W_{+2}^{0}|\tilde{t}_{\sigma}|^{-\gamma -1},  \nonumber 
 \end{eqnarray}
where $J_{1}$ and $\tilde{t}_{\sigma}$ are defined in (\ref{eq:3.27}) and (\ref{eq:4.73}), respectively.

\subsection{Spectator Phase Boundary: Isotherms below $T_{e}$}
\label{sec4.f}
The spectator-phase boundary, $g_{\sigma}(t,h)$, below the endpoint temperature can be obtained as in the previous section by using the expansion (\ref{eq:2.21}) for the scaling function $W_{-}(y,y_{4},\cdots)$ in (\ref{eq:4.33}). The $|y|$ factors in (\ref{eq:2.21}) yield the two branches of the phase boundary, $g_{\sigma}(t,h)$: see Figs.~\ref{fig2}(a) and~\ref{fig5}(a).

In the {\em symmetric case}, combining the results in the previous section with the expansion (\ref{eq:2.21}) yields
 \begin{equation}
  D_{1}g + D_{2}t + \cdots = -QW_{-0}^{0}|\tilde{t}|^{2-\alpha} - QW_{-1}^{0}U|\tilde{t}|^{\beta}|\tilde{h}| - QW_{-2}^{0}U^{2}|\tilde{t}|^{-\gamma}\tilde{h}^{2} + \cdots.   \label{eq:4.75}
 \end{equation}
Solving this for $g$ with the aid of (\ref{eq:4.65}) provides the result
 \begin{equation}
  g_{\sigma}(t;\tilde{h}) = - g_{\sigma,0}^{-} - g_{\sigma,1}^{-}|t|^{2-\alpha} \mp g_{\sigma,2}^{-}|t|^{\beta}\tilde{h} - g_{\sigma,3}^{-}|t|^{-\gamma}\tilde{h}^{2} + \cdots,   \label{eq:4.76}
 \end{equation}
where the upper (lower) sign corresponds to $\tilde{h} > 0$ ($<0$) while the coefficients are
 \begin{eqnarray}
  g_{\sigma,0}^{-} & = & D_{2}/D_{1}, \hspace{.2in} g_{\sigma,1}^{-} = QD_{1}^{-3+\alpha}W_{-0}^{0}|D_{1}-q_{1}D_{2}|^{2-\alpha},     \nonumber \\
  g_{\sigma,2}^{-} & = & QD_{1}^{-1-\beta}W_{-1}^{0}U|D_{1}-q_{1}D_{2}|^{\beta}, \hspace{.2in} g_{\sigma,3}^{-} = QD_{1}^{\gamma -1}W_{-2}^{0}U^{2}|D_{1}-q_{1}D_{2}|^{-\gamma}.  \label{eq:4.77}
 \end{eqnarray}
Notice that the linear term in $\tilde{h}$ vanishes as $T \rightarrow T_{e}-$, while the coefficient of the $\tilde{h}^{2}$ term diverges. In terms of $h$ we obtain the same leading $t$-dependent coefficients for $g_{\sigma}(t;h)$.

Finally, in the {\em nonsymmetric case} we obtain the equation
 \begin{eqnarray}
  (D_{1}-r_{0}D_{3})g & + & (D_{2}-r_{-1}D_{3})t + D_{3}\tilde{h} + \cdots  \nonumber \\
  & = & -QW_{-0}^{0}|\tilde{t}|^{2-\alpha} - QW_{-1}^{0}U|\tilde{t}|^{\beta}|\tilde{h}| - QW_{-2}^{0}U^{2}|\tilde{t}|^{-\gamma}\tilde{h}^{2} + \cdots.   \label{eq:4.78}
 \end{eqnarray}
By using (\ref{eq:4.65}), we can solve this for $g$ to obtain
 \begin{equation}
  g_{\sigma}(t;\tilde{h}) = -g_{\sigma,0}^{-}t - g_{\sigma,1}^{-}|t|^{2-\alpha} - (J_{1} \pm g_{\sigma,2}^{-}|t|^{\beta})\tilde{h} - g_{\sigma,3}^{-}|t|^{-\gamma}\tilde{h}^{2} + \cdots,  \label{eq:4.79}
 \end{equation}
where, again, the upper (lower) sign corresponds to $\tilde{h} > 0$ ($<0$), while the coefficients are
 \begin{eqnarray}
  g_{\sigma,0}^{-} & = & \frac{(D_{2}-r_{-1}D_{3})}{(D_{1}-r_{0}D_{3})}, \hspace{.2in}  g_{\sigma,1}^{-} = \frac{QW_{-0}^{0}}{(D_{1}-r_{0}D_{3})}|\tilde{t}_{\sigma}|^{2-\alpha}, \nonumber \\
  g_{\sigma,2}^{-} & = & \frac{QW_{-1}^{0}U}{(D_{1}-r_{0}D_{3})}|\tilde{t}_{\sigma}|^{\beta}, \hspace{.2in}  g_{\sigma,3}^{-} = \frac{QW_{-2}^{0}U^{2}}{(D_{1}-r_{0}D_{3})}|\tilde{t}_{\sigma}|^{-\gamma}. \label{eq:4.80}
 \end{eqnarray}
Notice that the linear term in $\tilde{h}$ does not vanish in the {\em nonsymmetric case}, but the slopes of two branches approach the same value as $T \rightarrow T_{e}-$. The coefficient of the quadratic term in $\tilde{h}$ diverges as the endpoint temperature is approached from below. As before the result (\ref{eq:4.79}) can be expressed in terms of $h$ by using (\ref{eq:4.74}).

\vspace{-.2in}
\section{Conclusions}
\label{sec5}
\vspace{-.15in}
In summary, following earlier studies \cite{ref8,ref9} stimulated by Widom, \cite{ref9} we have investigated the singular shapes of the various isothermal binodals, or two-phase coexistence curves, in the density plane near a critical endpoint. However, whereas the previous studies assumed classical or van der Waals expressions for the critical thermodynamics, our work is based on {\em nonclassical} phenomenological scaling postulates set out, in Sec.~\ref{sec2}, in a general form encompassing a spectrum of correction-to-scaling variables. Four types of critical endpoints were distinguished and examined in detail: {\em nonsymmetric}, labeled {\bf NA} or {\bf NB} depending on whether the lambda line $T_{\lambda}(g)$, which terminates at the endpoint $(g_{e},T_{e})$, lies, {\bf A}, below $T = T_{e}$ (as in Fig.~\ref{fig1}) or, {\bf B}, runs above (as in Fig.~\ref{fig4}); and {\em symmetric}, labeled, correspondingly, {\bf SA} and {\bf SB}: see Fig.~\ref{fig4}. At the endpoints, the lambda-line binodals ${\cal B}_{e}^{\lambda +}$ and ${\cal B}_{e}^{\lambda -}$ --- see Fig.~\ref{fig6}(b) --- were found to be singular with a leading ``renormalized'' exponent $(1-\alpha)/\beta$ and subdominant singular correction exponents. The symmetric $\lambda$ binodals are displayed in (\ref{eq:3.10}) [with explicit amplitude expressions recorded in (\ref{eq:4.27})-(\ref{eq:4.28})]; the nonsymmetric $\lambda$ binodals are presented in (\ref{eq:3.17}). 

Second, the {\em noncritical} or spectator-phase endpoint binodals ${\cal B}_{e}^{\alpha +}$ and ${\cal B}_{e}^{\alpha -}$ --- see Figs.~\ref{fig3}(b) and~\ref{fig6}(b) --- were found to be singular with a leading exponent $(\delta +1)/\delta$ (as conjectured by Widom \cite{ref9}): the symmetric binodals are given in (\ref{eq:3.36}) with the closely spaced sequence of correction exponents listed in (\ref{eq:3.38}). The nonsymmetric binodals are similar but more complicated: see (\ref{eq:3.44})-(\ref{eq:3.49}). The endpoint binodals ${\cal B}_{e}^{\beta}$ and ${\cal B}_{e}^{\gamma}$ which limit the critical phases --- see Figs.~\ref{fig3}(c) and~\ref{fig6}(c) --- have also been studied and found to have the same leading exponent, $(1-\alpha)/\beta$, as the lambda-line binodals: the symmetric forms are given in (\ref{eq:3.52}), the nonsymmetric expressions in (\ref{eq:3.56}). 

In addition, {\em above} the endpoint temperature the binodals separating the spectator-phase from the near-critical phase --- see ${\cal B}^{\alpha}$ and ${\cal B}^{\beta\gamma}$ in Figs.~\ref{fig3}(c) and~\ref{fig6}(c) --- were studied. They are analytic, but their slopes and curvatures develop singularities as $T \rightarrow T_{e}+$. The spectator-phase binodal is given in (\ref{eq:3.58}) and (\ref{eq:3.59}): its curvature diverges like $(T - T_{e})^{-\gamma}$ when the critical endpoint is approached. The conjugate, near-critical-phase binodal is described by (\ref{eq:3.64}) and (\ref{eq:3.68}); but in this case both the slope and the curvature {\em vanish}, although in a singular fashion, on approaching the endpoint. 

Finally, the binodals that approach the three-phase region below the endpoint temperature have been considered. The spectator-phase binodals ${\cal B}_{<}^{\alpha -}$ and ${\cal B}_{<}^{\alpha +}$ --- see Figs.~\ref{fig3}(a) and~\ref{fig6}(a) --- are presented in (\ref{eq:3.69}) and (\ref{eq:3.70}): as above the endpoint, their curvatures both diverge when $T \rightarrow T_{e}$.

Our analysis has utilized certain essential convexity or thermodynamic stability properties at and near a critical endpoint and, for Ising-type criticality, also invoked a specific positivity of a scaling function expansion coefficient: see the discussion after Eqns (\ref{eq:2.20}) and (\ref{eq:2.24}). These features are taken up in Ref. 17. 

\vspace{-.1in}
\acknowledgements
\vspace{-.1in}
We are grateful to Professor Ben Widom whose original questions stimulated our investigations. The support of the National Science Foundation (through Grant No. CHE 99-81772 and earlier grants) has been appreciated. MCB acknowledges support from the Conselho Nacional de Desenvolvimento Cientifico e Technologico (CNPq) and from Financiadora de Estudos e Projectos (Finep).

\begin{figure}
\caption{ The thermodynamic field space $(g, T, h)$ showing a {\em non}symmetric ({\bf N}) critical endpoint, $E$, at the meet of a $\lambda$ line, marking the edge of a phase boundary surface $\rho$ on which phases $\beta$ and $\gamma$ can coexist, and a phase boundary surface $\sigma$ limiting the spectator phase $\alpha$. The triple line $\tau$, on which phases $\alpha$, $\beta$ and $\gamma$ may coexist, extends above $T=T_{e}$ into the dot-dash line $\tilde{\tau}$ which is the intersection of $\sigma$ with the extended phase boundary $\tilde{\rho}$ (not shown). Note, as discussed below, that the $\lambda$ line shown here slopes downwards towards the $\alpha$ phase as $T$ rises, so representing case {\bf A}.}
\label{fig1}
\vspace{.3in}
\caption{ Isothermal sections of the {\bf NA} endpoint phase diagram in the field space in Fig. 1 for (a) $T < T_{e}$, (b) $T = T_{e}$, and (c) $T > T_{e}$. For $T \leq T_{e}$ the phase boundary $\sigma$ breaks into two pieces: $\sigma_{+}$ separating phases $\alpha$ and $\beta$, and $\sigma_{-}$ separating $\alpha$ and $\gamma$. The dotted curve represents the locus $\tilde{h}(g,T,h) = 0$ [see Sec. II] which coincides with the surface $\rho$ in Fig. 1 and defines its extension $\tilde{\rho}$ and, hence, the extended triple line $\tilde{\tau}$.}
\label{fig2}
\vspace{.3in}
\caption{ Isothermal density-density (or composition) diagrams for (a) $T < T_{e}$, (b) $T = T_{e}$, and (c) $T > T_{e}$ for an {\bf NA} endpoint showing single-phase regions $\alpha$, $\beta$, $\gamma$, and $\beta\gamma$, two-phase regions ruled by tie-lines connecting coexisting phases, and a three-phase triangle (dotted) in which coexist phases corresponding to the vertices $\tau\alpha$, $\tau\beta$, and $\tau\gamma$. The various analytically distinct binodals are labeled ${\cal B}_{<}^{\alpha -}$, ${\cal B}_{e}^{\gamma}$, $\cdots$, where the superscript indicates the phase bounded by the binodal while the subscript serves (as needed) to specify the temperature, $T \lessgtr T_{e}$. The same notations apply to a symmetric {\bf SA} endpoint. At $T = T_{e}$ the endpoint tie-line $E^{\alpha}E^{\lambda}$ defines the $\tilde{m}$ or $m = 0$ axis, shown dashed, where $m$ and $\tilde{m}$ are fixed linear combinations of the densities $\rho_{1}$ and $\rho_{2}$ (see Sec. III); the $m$ and $\tilde{m}$ axes on the plots in (a) and (c) have been omitted for the sake of clarity but are useful to understand the motion of the various features as $T$ passes through $T_{e}$.}
\label{fig3}
\vspace{.3in}
\caption{ Thermodynamic field space illustrating a {\em sym}metric critical endpoint, {\bf SB}, for case {\bf B} in which the $\lambda$ line slopes upwards away from the spectator phase $\alpha$ as $T$ increases. Beyond these differences, the phases, phase boundaries, etc., correspond precisely to those in Fig. 1.}
\label{fig4}
\vspace{.3in}
\caption{ Isothermal sections of the {\bf SB} endpoint phase diagram in Fig. 4 for (a) $T < T_{e}$, (b) $T = T_{e}$, and (c) $T > T_{e}$. Compare with Fig. 2 and note that for $T > T_{e}$ the $\lambda$ point and its phase boundary $\rho$ are disconnected from the boundary $\sigma$.}
\label{fig5}
\vspace{.3in}
\caption{ Isothermal density-density diagrams for (a) $T < T_{e}$, (b) $T = T_{e}$, and (c) $T > T_{e}$ for the {\bf SB} type of critical endpoint shown in Figs. 4 and 5. Compare the dispositions of the binodals with those shown in Fig. 3 and note the augmented labelling notation.}
\label{fig6}
\end{figure}


\begin{thebibliography}{99}
\bibitem{ref11} Note that $d = 4$ is a marginal dimensionality for simple critical behavior; thus this reduction as $d \rightarrow 4$ is only formal. In fact, one should anticipate logarithmic factors accompanying the $4/3$ power of the noncritical binodal when $d = 4$.
\bibitem{ref16} In renormalization group language we are thus supposing that no ``resonances'' of eigenexponents arise, at least to the orders of interest. Such resonances may lead to logarithmic factors. However, for most applications in $d = 3$ dimensions no such difficulties should be anticipated.
\bibitem{mixing} It can be seen that pressure-mixing will not alter the leading behavior of the various binodals, etc., but will enter the correction terms in various thermodynamic quantities.
\bibitem{ref17} Note that if $L_{\sigma}$ and $L_{\rho}$ should be infinite we may merely interchange the labels of the $g$ and $h$ fields. Similarly, to place the $\beta$ phase within the positive $(m, \tilde{m})$ quadrant the signs ascribed to the fields $g$ and $h$ may be appropriately adjusted.
\end{thebibliography}
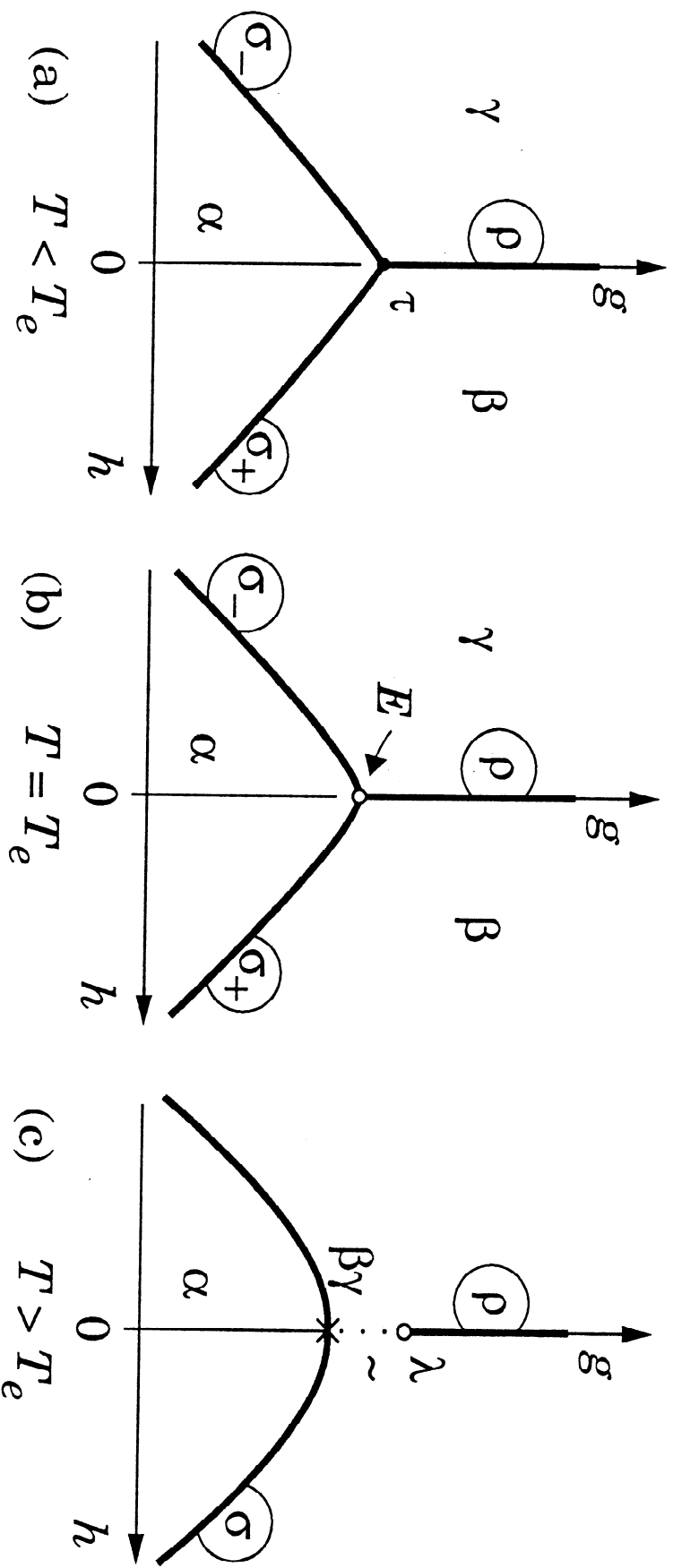
\end{document}